\documentclass{aa}

\usepackage[varg]{txfonts}
\usepackage{adjustbox,lipsum}

\usepackage{graphicx}
\usepackage{amsmath}
\usepackage[config, labelfont={rm,bf}, font=footnotesize, singlelinecheck=false]{caption, subfig}

\title{No evidence for small disk-like bulges in a sample of late-type spirals}

\author{L.~Costantin\inst{1}\thanks{luca.costantin@studenti.unipd.it}
\and J.~M\'endez-Abreu\inst{2}
\and E.~M.~Corsini\inst{1, 3}
\and L.~Morelli\inst{1, 3}
\and J.~A.~L.~Aguerri\inst{4, 5}
\and E.~Dalla~Bont\`a\inst{1, 3}
\and A.~Pizzella\inst{1, 3}
}

\institute {Dipartimento di Fisica e Astronomia `G. Galilei', Universit\`a di Padova, vicolo dell'Osservatorio 3, I-35122 Padova, Italy
\and School of Physics and Astronomy, University of St. Andrews, SUPA, North Haugh, KY16 9SS St. Andrews, UK
\and INAF - Osservatorio Astronomico di Padova, vicolo dell'Osservatorio 5, I-35122 Padova, Italy
\and Instituto de Astrof\'isica de Canarias, Calle V\'ia L\'actea s/n, E-38200 La Laguna, Tenerife, Spain
\and Departamento de Astrof\'isica, Universidad de La Laguna, Calle Astrof\'isico Francisco S\'anchez s/n, E-38205 La Laguna, Tenerife, Spain
}

\begin{document}

\abstract 
{About $20\%$ of low-redshift galaxies are late-type spirals with a small
or no bulge component. Although they are the simplest disk galaxies
in terms of structure and dynamics, the role of the different
physical processes driving their formation and evolution is not yet
fully understood.}
{We investigated whether small bulges of late-type spirals follow the same scaling
relations traced by ellipticals and large bulges and 
if they are disk-like or classical bulges.}
{We derived the photometric and kinematic properties of 9 nearby
late-type spirals. To this aim, we analyzed the surface-brightness
distribution from the \emph{i}-band images of the Sloan Digital Sky
Survey and obtained the structural parameters of the galaxies from a
two-dimensional photometric decomposition. We found a bulge component
in 7 galaxies of the sample, while the remaining 2 resulted to be pure disk
galaxies. We measured the line-of-sight stellar velocity distribution
within the bulge effective radius from the long-slit spectra taken
with high spectral resolution at the Telescopio Nazionale Galileo. We
used the photometric and kinematic properties of the sample bulges to study their location in the
Fundamental Plane, Kormendy, and Faber-Jackson relations 
defined for ellipticals and large bulges.}
{We found that our sample bulges satisfy some of the photometric and
kinematic prescriptions for being considered disk-like bulges such as
small sizes and masses with nearly exponential light profiles, small
bulge-to-total luminosity ratios, low stellar velocity dispersions,
and ongoing star formation. However, each of them follows the same scaling
relations of ellipticals, massive bulges, and compact early-type
galaxies so they cannot be classified as disk-like systems.}
{We found a single population of galaxy spheroids that follow the
same scaling relations, where the mass seems to lead to a smooth
transition in the photometric and kinematic properties from less
massive bulges to more massive bulges and ellipticals.}


\maketitle



\section{Introduction \label{sec:introduction}}

The question of whether the morphology of galaxies is mostly imprinted by the
initial conditions of their formation or by the processes driving
their evolution is still open. The Hubble scheme of morphological
classification tells us that spirals range from early to late types
mainly according to the bulge-to-disk luminosity ratio  \citep[see][for a review]{Buta2015}. 
Historically, this classification was done by the visual inspection of galaxies on
photographic plates. With the progress of imaging technology, new
methods to quantify the prominence of the bulge component were
developed. Nowadays, the description of spirals according to the bulge
size is usually performed by applying a photometric decomposition 
\citep[e.g.,][]{Simard1998, Peng2002,
deSouza2004, Mendez2008a, Erwin2015},
which also provides a full description of the bulge light
distribution. In this context, the bulge is photometrically defined as
the extra component dominating the galaxy light above the
surface-brightness profile of the disk extrapolated in the inner
regions of the galaxy. The bulge surface brightness is usually fitted
with a S\'ersic law spanning a large range of profile shapes, whereas
the disk surface brightness is usually fitted with an exponential law 
\citep{Andredakis1998, Prieto2001, Aguerri2005, Mendez2016}.

The current paradigm separates bulges into two main categories, namely
classical and disk-like bulges, according to their photometric,
kinematic, and stellar population properties which result from their
different assembly histories \citep{Athanassoula2005, Fisher2008, Laurikainen2016}.

In this picture, classical bulges are thought to be formed by rapid
gravitational collapse of protogalactic gas clouds \citep{Eggen1962}
or by accretion during major \citep{Kauffmann1996} and minor merging
events \citep{Aguerri2001, ElicheMoral2006}. Moreover, they could be
formed as a consequence of primordial clump instabilities at high-$z$
sinking rapidly toward the center on account of dynamical friction
\citep{Noguchi1999, Dekel2009, Bournaud2016}. Classical bulges are thought to be
similar to low-luminosity ellipticals, being characterized by rounder
shapes than their surrounding disks, power-law profiles, and old stellar
populations \citep[see][for a review]{Kormendy2004}.
Most of them are consistent with isotropic oblate rotators (but see \citealt{Mendez2016b},
for different findings) and follow the Fundamental Plane (FPR), Kormendy
(KR) and Faber-Jackson (FJR) relations \citep{Bender1992,
Aguerri2005}.

As their name suggests, disk-like bulges are reminiscent of
disks. Their formation involves secular processes of redistribution of
gas and stars driven by bars or environmental phenomena, including the
accretion of low-density satellites \citep{ElicheMoral2011}. 
They have ellipticities similar to those of their
disks, nearly-exponential luminosity profiles, and young stellar
populations \citep[e.g.,][]{Fisher2016}. They are rotationally-supported
 oblate spheroids \citep{Mendez2010} and tend to be low-$\sigma$ outliers in the FJR
\citep{Kormendy2016}.

More than $50\%$ of edge-on galaxies, including our Milky Way,
show in their central parts a thick component swelling out of the disk
plane, which has the shape of a box or a peanut \citep{Lutticke2000a,
Lutticke2000b}.  Although these structures are usually called
boxy/peanut (B/P) bulges, this terminology could be misleading because
they are just part of edge-on bars and are related to their
secular evolution \citep{Combes1981, Chung2004}.  In fact, the
fraction of barred galaxies is consistent with that of edge-on
galaxies with B/P bulges \citep{Eskridge2000}. Moreover, numerical
simulations of stellar orbits confirmed that B/P bulges are connected
to bars \citep{Bureau2005, Athanassoula2013} and photometric and 
kinematic observations showed that B/P bulges
share the same properties of bars \citep{Mendez2008b, Erwin2013}. The
whole picture is further complicated by the co-existence 
of classical and disk-like bulges in the same galaxy
\citep{Gadotti2009, Kormendy2010b, Erwin2015b} and their combination
with B/P structures in barred galaxies \citep{deLorenzo2012, Mendez2014}.

The population of late-type spirals comprises about $20\%$ of the
galaxies at low redshift \citep{Nair2010} and is composed by disks
with a small or no bulge component. Late-type spirals are
characterized by the presence of large amounts of dust, on-going star
formation, and small-scale substructures both in the nucleus and along
the arms \citep{Carollo1997}.
Moreover, some of them show a light excess and a peculiar kinematics
in the inner hundred parsecs due to the presence of a nuclear cluster
\citep{Boker2002, Seth2006}, a nuclear stellar disk
\citep{Morelli2010, Corsini2012}, or a nuclear bar \citep{Erwin2004,
deLorenzo2013}. Thus, measuring and interpreting the properties of
their bulges is particularly challenging. The extremely late-type
spirals located at the end of the Hubble morphological sequence do not
actually have a bulge component and are pure disk galaxies \citep{Boker2002,
Boker2003}.

Although bulgeless spirals are the simplest disk galaxies in terms of
structure and dynamics, the role of the different physical processes
driving their formation and evolution is not yet fully
understood. Indeed, hierarchical clustering scenarios are successful
in explaining the assembly and growth of massive ellipticals, but do
not fully account for the building of disk-dominated galaxies with
little or no bulge \citep{Kautsch2006}. Despite the fact that
\citet{Hopkins2009} showed that the gas content in mergers plays a
significant role in the efficiency of disk destruction and enables
disks to survive, numerical simulations still have problems to reproduce the 
observed fraction of extremely-late spirals in the local universe
\citep{Weinzirl2009, Kormendy2010, Zavala2012}.

The proper way to classify galactic bulges
would be distinguishing between their different formation scenarios
which led to different observed properties. Reconstructing the
formation mechanism from observations is a hard task, but the way in
which scaling relations are satisfied by galaxies reflects the general
dynamical principles of how baryons settle into dark matter potential wells.
Previous works reported some hints about differences between less
and more massive bulges suggesting different formation scenarios
and/or evolution pathways \citep{Graham2003, Mendez2008a}. However,
since observational studies on bulges were mostly focused on the
prominent ones, further efforts are required to explore the
photometric and kinematic properties of small and low-mass bulges in
nearby late-type spirals. This is the aim of this paper where we
extend the scaling relations of bulges all the way down to the lowest
mass regime studied so far. 

The paper is organized as follows. We present the galaxy sample in
Sect. \ref{sec:sample}. We explain the acquisition, reduction, and
analysis of the photometric and spectroscopic data in
Sect. \ref{sec:photometry} and \ref{sec:spectroscopy}, respectively.
We investigate whether our small bulges follow the same scaling
relations traced by ellipticals and large bulges in
Sect. \ref{sec:scaling_relations}.  We discuss the results and give
our conclusions in Sect. \ref{sec:discussion} and
\ref{sec:conclusions}, respectively. We adopt $H_0 = 75$ km s$^{-1}$
Mpc$^{-1}$, $\Omega_{\rm M} = 0.3$, and $\Omega_{\Lambda} = 0.7$ as
cosmological parameters throughout this work.


\section{Sample selection \label{sec:sample}}

\begin{table*}[t]
\centering
\caption{Properties of the sample galaxies.}
\begin{adjustbox}{max width=\linewidth}
\begin{tabular}{ccccccccccccc}
\hline
\hline
\multicolumn{2}{c}{Galaxy}      & \multicolumn{2}{c}{Coordinates}              & \multicolumn{2}{c}{Type}	& $B_{\rm T}$ & $i$	 & $z$ & $D$  & Scale              & $D_{25}\times d_{25}$ & $M_{B_T}^{0}$ \\
SDSS name		   & Alternative name  & RA [h m s] & DEC [deg arcmin arcsec] &   &   & [mag]	    & [$^{\circ}$] &    & [Mpc] & [pc arcsec$^{-1}$] & [arcsec]		& [mag]        \\
(1)					       & (2)	& (3)	    & (4)	 & (5) & (6)  & (7)		   & (8)		& (9)     & (10) & (11)   & (12) & (13)  \\
\hline
SDSS~J104054.43$+$143202.4 &	...	       & 10~40~54.43 & $+$14~32~02.4 & ...	& Sa-Sb (0.48) & 16.00 & 22.4 & 0.021 & 81.4 &	395 & $21\times20$ & $-$19.0\\
SDSS~J112139.74$+$112924.6 &	...	       & 11~21~39.74 & $+$11~29~24.6 & ...	& Sc-Sd (0.47) & 16.77	& 20.3 & 0.020 & 77.6 &	378 & $19\times17$ & $-$18.1\\
SDSS~J113642.30$+$545045.7 &	IC~2943	        & 11~36~42.30 & $+$54~50~45.7 & Sa	& Sc-Sd (0.49) & 15.18	& 39.5 & 0.019 & 75.8 &	368 & $24\times19$ & $-$19.6\\
SDSS~J115243.42$+$014428.0 &	UGC~6854        & 11~52~43.42 & $+$01~44~28.0 & SBbc	& Sa-Sb (0.49) & 14.52	& 39.6 & 0.020 & 79.6 &	386 & $56\times43$ & $-$20.2\\
SDSS~J133253.11$-$011531.1 &	PGC~47684       & 13~32~53.11 & $-$01~15~31.1 & S?&  Sa-Sb (0.35) & 15.70	& 40.6 & 0.012 & 47.1 &	228 & $29\times22$ & $-$18.3\\
SDSS~J143227.42$+$272538.7 &	IC~4452	       & 14~32~27.42 & $+$27~25~38.7 & Sa	& Sc-Sd (0.62) & 14.92	& 20.6 & 0.014 & 55.9 &	271 & $29\times27$ & $-$19.2\\
SDSS~J144425.40$+$415140.6 &	PGC~2188136    & 14~44~25.40 & $+$41~51~40.6 & ...& S0 (0.36) & 15.79	& 36.0 & 0.018 & 69.0 &	335 & $39\times31$ & $-$18.6\\
SDSS~J160324.17$+$205328.4 &	NGC~6035       & 16~03~24.17 & $+$20~53~28.4 & Sc	& Sc-Sd (0.64) & 14.20	& 29.0 & 0.016 & 62.2 &	301 & $59\times52$ & $-$20.4\\
SDSS~J170128.21$+$634128.0 &	IC~1241	        & 17~01~28.21 & $+$63~41~28.0 & Sc	& Sc-Sd (0.63) & 14.37	& 38.2 & 0.016 & 63.4 &	308 & $70\times56$ & $-$20.0\\
\hline  
\end{tabular}
\end{adjustbox}
\tablefoot{(1) Full name of the galaxy according to SDSS. (2) Alternative
name. (3), (4) Right ascension and declination (J2000.0). (5) Morphological classification from HyperLeda
\citep{Makarov2014}. (6) Morphological classification and corresponding probability 
from \citet{HuertasCompany2011}.
(7) Apparent total blue magnitude from
HyperLeda. (8) Inclination from HyperLeda. (9) Spectroscopic
redshift from SDSS DR6. (10) Angular
diameter distance. (11) Conversion factor from arcsec to parsec. (12) Major
and minor diameters of the isophote at a surface brightness level of
$\mu_B = 25$ mag~arcsec$^{-2}$ from the HyperLeda. (13) Absolute
total blue magnitude corrected for inclination and extinction from
the HyperLeda.}
\label{tab:sample}
\end{table*}

\begin{figure}[t!]
\centering
\resizebox{\hsize}{!}{\includegraphics{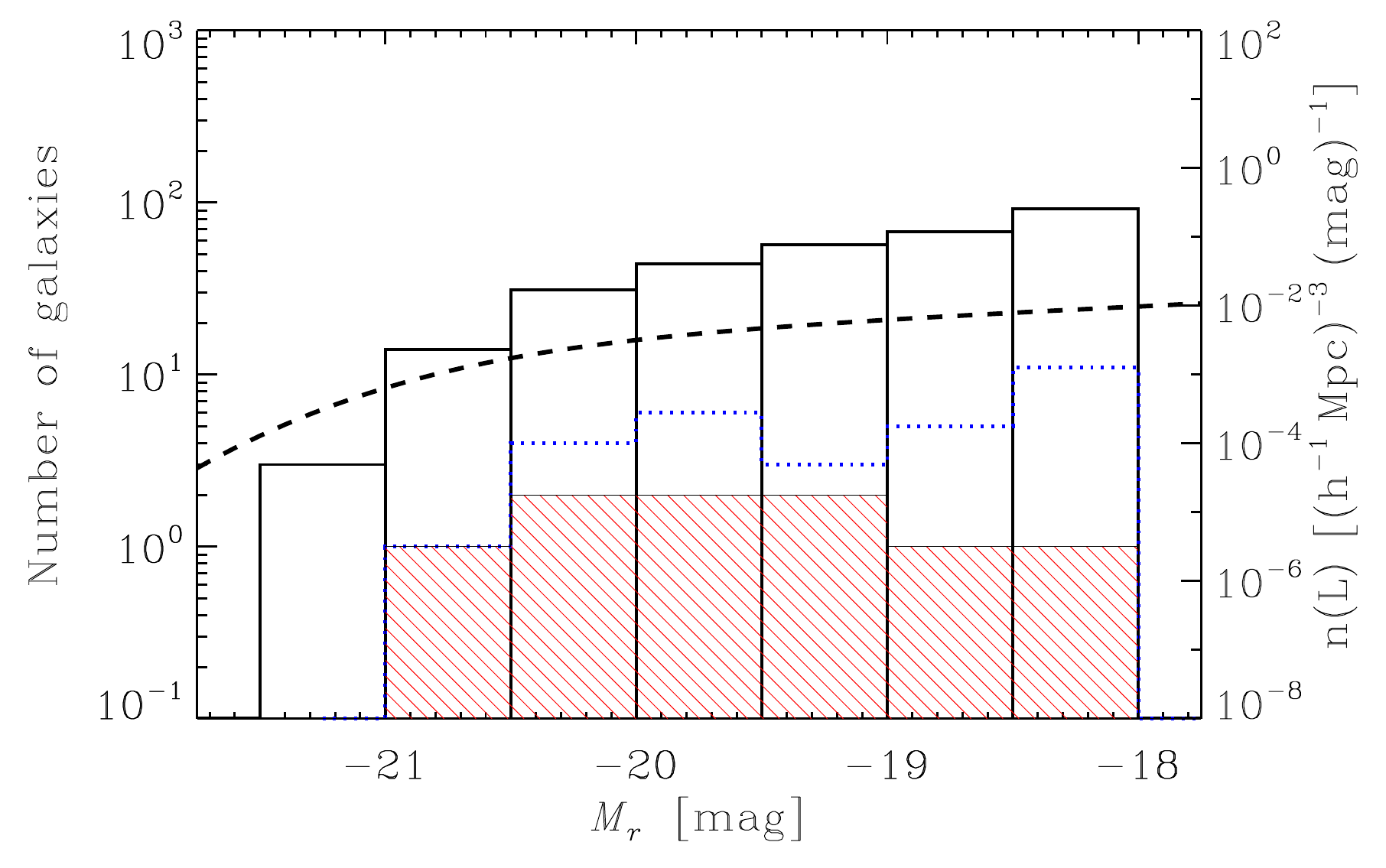}}
\caption{Distribution of the absolute \emph{r}-band magnitude of the
volume-limited sample (309 galaxies, black solid line),
representative sample of late-type spirals (30 galaxies, blue dotted
line), and observed sample (9 galaxies, red filled histogram).
The dashed line corresponds to the $r$-band luminosity 
function of SDSS galaxies by \citet{Tempel2011}.
}
\label{fig:sample}
\end{figure}

We selected a volume-limited sample of 309 disk galaxies located within a
radius of 100 Mpc and with $M_r \le -18$ mag using the Sloan Digital
Sky Survey (SDSS) Data Release 6 (DR6) catalogue
\citep{Adelman2008}. Since small bulges
are preferentially located in late-type spirals, we considered only
galaxies with a light concentration index $C = R_{90}/R_{50} < 2.5$
where $R_{50}$ and $R_{90}$ are the radii enclosing $50\%$ and $90\%$
of the total galaxy luminosity, respectively. It corresponds to a
S\'ersic index $n \le 1.5$ \citep{Conselice2003, Graham2005}. 
Since galaxies are randomly
oriented, it is possible to derive their properties by taking into
account only the objects within a inclination interval. Therefore, we
restricted to low inclination galaxies with $i < 45^{\circ}$ in order
to perform a successful photometric decomposition. Finally, we chose
the 30 closest objects to be representative of the low-inclined
late-type galaxies, but we observed only 9 of them due to time and
weather constraints during the spectroscopic observations.

The main properties of the sample galaxies are listed in
Table~\ref{tab:sample}, while the distributions of the \emph{r}-band
magnitudes of the volume-limited, representative, and final samples of
galaxies are shown in Fig.~\ref{fig:sample}.
Our sample galaxies are all late-type
spirals (Fig.~\ref{fig:decomposition}), despite their morphological 
classification in \citet{HuertasCompany2011} and \citet{Makarov2014}.


\section{Surface photometry \label{sec:photometry}}

\subsection{Data reduction}

\begin{table*}[t]
\caption{Characteristics of the \emph{i}-band SDSS images of the
sample galaxies.}
\centering
\begin{tabular}{ccccccc}
\hline
\hline
Galaxy & Gain		     & RON	& Sky   & FWHM     & $\beta$\\
       & [$e^{-}$~ADU$^{-1}$] & [$e^{-}$]	& [ADU] & [arcsec] &	    \\
(1)    & (2)		     & (3)	& (4)	& (5)      & (6)    \\
\hline
SDSSJ1040 & 4.8 & 10.7 & $212\pm6$ & 1.0 & 3.0\\
SDSSJ1121 & 6.6 & 16.4	& $177\pm5$ & 1.1 & 3.4\\
SDSSJ1136 & 4.9 & 13.4	& $230\pm6$ & 1.0 & 4.8\\
SDSSJ1152 & 6.6 & 16.4	& $121\pm4$ & 1.0 & 2.7\\
SDSSJ1332 & 4.8 & 10.7	& $184\pm5$ & 1.3 & 3.8\\
SDSSJ1432 & 6.6 & 16.4	& $150\pm5$ & 1.0 & 5.6\\
SDSSJ1444 & 4.9 & 13.4	& $199\pm5$ & 0.7 & 3.6\\
SDSSJ1603 & 6.6 & 16.4	& $128\pm4$ & 0.9 & 4.7\\
SDSSJ1701 & 4.9 & 10.4	& $119\pm5$ & 1.0 & 4.1\\
\hline  
\end{tabular}
\tablefoot{(1) Short name of the galaxy according to SDSS. (2) Image gain
provided by SDSS. (3) Image readout noise provided by SDSS. (4)
Measured sky level and corresponding standard deviation. (5), (6)
FWHM and $\beta$ parameter measured for the circular Moffat PSF.}
\label{tab:photometry}
\end{table*}

We retrieved the flux-calibrated \emph{i}-band images of the sample
galaxies from the SDSS DR9 \citep{Ahn2012}. The choice of \emph{i}
band ensured us a sufficient spatial resolution ($\rm FWHM \simeq 1$
arcsec) and depth (out to $\mu_i \simeq 25$ mag~arcsec$^{-2}$), and
minimized the dust effects with respect to the other SDSS passbands in
order to resolve the bulge component with an accurate photometric
decomposition of the galaxy surface-brightness
distribution. 

The sky level provided by SDSS for each image consists of a global
estimate across the field of view after masking the most luminous
sources \citep{Aihara2011}. It results to be unreliable in the analysis 
of the faintest outskirts of large galaxies like ours
\citep{Hyde2009, Mendez2016}. Therefore, we measured the sky level in the 
surroundings of each sample galaxy following the procedure proposed by
\citet{Pohlen2006} as applied in \citet{Corsini2016} and
\citet{Morelli2016}. We masked all the foreground stars,
companion and background galaxies, and spurious sources like residual
cosmic rays and bad pixels close to the galaxy and measured its
surface-brightness radial profile with the \texttt{ellipse} task in
IRAF\footnote{Image Reduction and Analysis Facility is distributed by
the National Optical Astronomy Observatory (NOAO), which is operated
by the Association of Universities for Research in Astronomy (AURA),
Inc. under cooperative agreement with the National Science
Foundation.} \citep{Jedrzejewski1987}.
First, we allowed to vary the center, ellipticity, and position angle
of the fitting ellipses. Then, we fitted again the isophotes with
\texttt{ellipse} adopting the center of the inner ellipses and the ellipticity
and position angle of the outer ones. Finally, we assumed the constant
value of the surface brightness measured at large radii, where the
galaxy contributed no light, as the sky level to be subtracted from
the image. We found that our estimates of the sky level are
systematically lower by $0.2\%$ than those given by SDSS. We
measured the standard deviation $\sigma_{\rm sky}$ of the background
in the sky-subtracted image by analyzing with the IRAF task \texttt{imexamine}
several regions where no sources were present. 

We trimmed the sky-subtracted images in order to reduce the computing
time to perform a reliable photometric decomposition. Each galaxy was
centered in a field of view of at least $50\times50$ pixels
corresponding to $20\times20$ arcsec$^2$. We ran \texttt{ellipse} on the trimmed images to derive
the radial profiles of the ellipse-averaged surface brightness $\mu$,
ellipticity $\epsilon$, and position angle $PA$ of the galaxy
isophotes in the \emph{i} band. We adopted these photometric profiles
as well as the mask images we built to estimate the sky level for the
photometric decomposition.

We modeled the point spread function (PSF) with a circular Moffat
profile \citep{Moffat1969, Trujillo2001}, which is a reliable approximation
for photometric decomposition of
nearby galaxies in SDSS images \citep{Mendez2016}, and
derived its parameters fitting  5 unsaturated
stars with the IRAF task \texttt{imexamine} before trimming the image (Table
\ref{tab:photometry}).

\subsection{Photometric decomposition}

We derived the structural parameters of the sample galaxies by
performing a two-dimensional photometric decomposition of their
surface brightness distribution using the GASP2D algorithm (GAlaxy
Surface Photometry 2 Dimensional Decomposition) developed by
\citet{Mendez2008a, Mendez2014} in IDL\footnote{Interactive Data
Language is distributed by ITT Visual Information Solutions. It is
available from \url{http://www.ittvis.com}}. GASP2D adopts a set of
analytical functions to model the light contribution of the galaxy
components. We assumed the surface brightness distribution
of the sample galaxies to be the sum of a bulge, disk, and bar
component. We did not consider any other additional component, such as
spiral arms, lenses or ovals.
When possible, we masked their corresponding regions in 
the galaxy images and excluded them from the fitting process.
In addition, we masked out as
much as possible the dust patches and lanes to recover a reliable
model of the surface-brightness distribution.

We used the S\'ersic law \citep{Sersic1968} to describe the surface
brightness of the bulge component
\begin{equation}
I_{\rm bulge}(x, y) = I_{\rm e} 10^{-b_n \left[ \left(\dfrac{r_{\rm bulge}}{r_{\rm e}} \right)^{1/n} - 1 \right]} ,
\end{equation}
where $r_{\rm e}$ is the effective radius, $I_{\rm e}$ is the surface
brightness at $r_{\rm e}$, $n$ is a shape parameter describing the
curvature of the surface brightness profile, and $b_n = 0.868 n -
0.142$ \citep{Caon1993}. We assumed the bulge isophotes to be
elliptical and centered on the galaxy center $(x_0,y_0)$, with
constant position angle $PA_{\rm bulge}$ and constant axial
ratio $q_{\rm bulge}$.  The radius $r_{\rm bulge}$ is given by
\begin{equation}
\begin{split}
r_{\rm bulge} & (x, y) = \biggl[ \left( - \left( x - x_0 \right) \sin PA_{\rm bulge} + 
  \left( y - y_0 \right) \cos PA_{\rm bulge} \right)^2 \\
            & + \left( \left( x - x_0 \right) \cos PA_{\rm bulge} + 
  \left( y - y_0 \right) \sin PA_{\rm bulge} \right)^2 / q_{\rm bulge}^2 
  \vphantom{\int_1^2} \biggl]^{1/2} \, .
\end{split}
\end{equation}

We adopted the exponential law \citep{Freeman1970} to describe the
surface brightness of the disk component
\begin{equation}
I_{\rm disk}(x, y) = I_0 e^{- \left(\dfrac{r_{\rm disk}}{h} \right)} ,
\end{equation}
where $I_0$ is the central surface brightness and $h$ is the
scalelength of the disk. We adopted elliptical isophotes for the
disk, which were centered on the galaxy center $(x_0,y_0)$ and had
constant position angle $PA_{\rm disk}$ and constant axial
ratio $q_{\rm disk}$.  The radius $r_{\rm disk}$ is given by
\begin{equation}
\begin{split}
r_{\rm disk} & (x, y) = \biggl[ \left( - \left( x - x_0 \right) \sin PA_{\rm disk} 
  + \left( y - y_0 \right) \cos PA_{\rm disk} \right)^2 \\
	   & + \left( \left( x - x_0 \right) \cos PA_{\rm disk} 
  + \left( y - y_0 \right) \sin PA_{\rm disk} \right)^2  / q_{\rm disk}^2 
  \vphantom{\int_1^2} \biggl]^{1/2} \, .
\end{split}
\end{equation}

The surface brightness of the bar was described using the projected
surface density of a three-dimensional Ferrers ellipsoid
\citep{Ferrers1877}
\begin{equation}
I_{\rm bar}(x, y) =
\begin{cases}
I_{0, \rm bar}  \left[ 1 - \left(\dfrac{r_{\rm bar}}{a_{\rm bar}} \right)^{2} \right]^{n_{\rm bar} + 0.5}        
  & \text{if $r_{\rm bar} \le a_{\rm bar}$} \\
0 & \text{if $r_{\rm bar} > a_{\rm bar}$}
\end{cases}
\, ,
\end{equation}
where $I_{0, \rm bar}$ represents the central surface brightness,
$a_{\rm bar}$ is the length, and $n_{\rm bar}$ is the shape parameter
of the bar. We assumed the isophotes of the bar to be generalized
ellipses \citep{Athanassoula1990} centered on the galaxy center
$(x_0,y_0)$, with constant position angle $PA_{\rm bar}$ and
constant axial ratio $q_{\rm bar}$.  The $r_{\rm bar}$
radius is defined as
\begin{equation}
\begin{split}
r_{\rm bar} & (x, y) = \biggl[ |\left( y - y_0 \right) \cos PA_{\rm bar}
- \left( x - x_0 \right) \sin PA_{\rm bar} |^c \\
	   & + |\left( \left( x - x_0 \right) \cos PA_{\rm bar} 
  + \left( y - y_0 \right) \sin PA_{\rm bar} \right)  / q_{\rm bar}|^c 
  \vphantom{\int_1^2} \biggl]^{1/c} \, ,
\end{split}
\end{equation}
where $c$ controls the shape of the bar isophotes. It serves the same purpose as
the $\cos{4\theta}$ Fourier coefficient which is usually adopted to
describe the boxiness/diskyness of the isophotes
\citep{Jedrzejewski1987, Bender1987}. A bar with pure elliptical
isophotes has $c = 2$. It is $c > 2$ if the isophotes are boxy and $c
< 2$ if they are disky. Including the bar in the multi-component
approach to the photometric decomposition is crucial to retrieve the
correct bulge parameters. In fact, both the S\'ersic index $n$ and
bulge-to-disk ratio $B/T$ can be overestimated if the bar is not
properly accounted for \citep{Aguerri2005, Gadotti2009, Mendez2014}.

We derived the best-fitting values of the structural parameters of the
galaxy components by comparing the model of the surface brightness
distribution $I_{\rm mod}$ with the observed photon counts of the
galaxy $I_{\rm obs}$ in each image pixel $(x,y)$ with a iterative
procedure of non-linear least-squares minimization based on a robust
Levenberg-Marquardt method \citep{More1980} using the IDL task MPFIT
\citep{Markwardt2009}. We convolved the model image with the circular
Moffat PSF measured on the galaxy image in order to deal with the
seeing effects. Each image pixel was weighted according to the
variance of its total observed photon counts due to the contribution
of both the galaxy and sky, in which the photon noise limitation and
readout noise of the detector were also taken into account.

We detected a bulge component in 7 out of 9 sample galaxies, whereas 
the remaining 2 (namely SDSSJ1040 and 
SDSSJ1121) were better fitted only with a disk component.
We detected a bar component in 4 out of 9 galaxies (namely 
SDSSJ1152, SDSSJ1444, SDSSJ1603, and SDSSJ1701).
The best-fitting structural parameters of the sample galaxies are
available in Table~\ref{tab:decomposition} while their GASP2D fits are
shown in Fig.~\ref{fig:decomposition}.

\subsection{Error budget\label{sec:err_bud}}

The formal errors obtained from the $\chi^2$ minimization procedure
are usually not representative of the real errors in the structural
parameters space \citep{Mendez2008a}.  Therefore, we estimated the
errors analyzing a sample of mock galaxies built through a series of
Monte Carlo (MC) simulations, in order to test the sensibility of the best-fitting
parameters and understand if any systematic error affects them.

Instead of randomly generating the mock galaxies, we created them
taking into account the correlations among the structural parameters
of bulge and disk. For this purpose we used the sample of nearby
galaxies of \citet{Gadotti2009}.  Firstly, we divided the 9 galaxies
of our sample in three subsamples, according to the number of their
components: (1) disk only, (2) bulge and disk, (3) bulge, disk, and
bar. This allowed us to define bins of magnitude in which we created
the corresponding sample of mock galaxies: 4 bins for subsamples (1) and (2) in
the magnitude range $-20.5 < M_i < -18.5$ mag and 4 bins for subsample
(3) in the magnitude range $-21.5 < M_i < -19.5$ mag. 
Each bin is composed of about $200$ mock galaxies. Then, we
generated a $B/T$ random value from a uniform distribution in the
range $[0, 0.45]$, and starting from this value we determined the
S\'ersic index $n$ from a normal distribution using the correlation
\begin{equation}
n = 5.73 \cdot B/T + 1.25
\label{eq:n_BT}
\end{equation}
we obtained by fitting the \citet{Gadotti2009} sample.  After that, we
produced a random value of $r_{\rm e}$ from a normal distribution and
determined a value of $h$ using the correlation
\begin{equation}
r_{\rm e}/h = 0.69 \cdot B/T + 0.19
\label{eq:rh_BT}
\end{equation}
we obtained from the sample of \citet{Gadotti2009}.  Unfortunately, we
could not use this sample to investigate the properties of the bar, so
we randomly generated the structural parameters of the bar in the same
intervals as those obtained from the photometric decomposition of our
sample.  The interval ranges explored for all the parameters were
$r_{\rm e} = [0.5, 4]$ arcsec, $n = [0.5, 4]$, $h = [1, 14]$ arcsec,
$a_{\rm bar} = [5, 25]$ arcsec, $n_{\rm bar} = [0, 7]$, and $c = [-2,
5]$.  Finally, we randomly generated the apparent flattening of the
bulge $q_{\rm bulge}$, disk $q_{\rm disk}$, and bar $q_{\rm bar}$ as
well as the values of the position angle of the bulge $PA_{\rm
bulge}$, disk $PA_{\rm disk}$, and bar $PA_{\rm bar}$
from uniform distributions, where no constraints were adopted.

We assumed the mock galaxies to be at a distance of 69 Mpc, which
corresponds to the median distance of our sample galaxies.  We chose
the pixel scale (0.396 arcsec pixel$^{-1}$), gain (6.6
$e^{-}$~ADU$^{-1}$), readout noise (16.4 $e^{-}$ rms), and size of the simulated
images ($400\times400$ pixels) to mimic the instrumental setup of the
photometric observations.  Moreover, we added a background level (170
ADU) and a Poissonian photon noise to yield in the simulated images a
signal-to-noise ratio ($S/N$) similar to that of the observed ones.

We analyzed the images of the mock galaxies as if they were real using
GASP2D. Thus, we evaluated the initial conditions in the fitting
procedure starting from the values of the generated parameters for
each model galaxy.  We estimated the errors on the fitted parameters
by comparing the input and measured values, assuming they were
normally distributed. For $I_{\rm e}$, $r_{\rm e}$, $n$, $I_{0}$, $h$,
$I_{0, \rm bar}$, $a_{\rm bar}$, $n_{\rm bar}$, and $c$ we adopted the
mean and standard deviation of the relative errors of the mock
galaxies as the systematic and statistical errors of the observed
galaxies, respectively. For $q_{\rm bulge}$, $q_{\rm disk}$, $q_{\rm
bar}$, $PA_{\rm bulge}$, $PA_{\rm disk}$, and
$PA_{\rm bar}$ we adopted the mean and standard deviation of
the absolute errors of the mock galaxies as the systematic and
statistical errors of the observed galaxies, respectively. 
The computed errors  $\sigma^2$ = ${\sigma_{\rm stat}^2 + \sigma_{\rm syst}^2}$ are provided in Table~\ref{tab:decomposition}, being the systematic errors
$\sigma_{\rm syst}$ negligible compared to the statistical errors $\sigma_{\rm stat}$.
In the analysis we did not consider the systematic
errors derived from the uncertainties on the estimates of the PSF FWHM
or sky level. Therefore, the errors reported in Table
\ref{tab:decomposition} could be slightly underestimated
\citep[see][for a discussion]{Mendez2008a}.

Finally, we investigated the accuracy in measuring the structural
parameters of the bulges of the sample galaxies in spite of their
small size and demonstrated that we obtained reliable values (Appendix
\ref{appendix:resolution}).

\begin{sidewaystable}[h!]
\centering
\caption{Structural parameters of the sample galaxies.}
\begin{adjustbox}{max width=\linewidth}
\begin{tabular}{ccccccccccccccccccc}
\hline
\hline
galaxy  				& 	$\mu_{\rm e}$		& $ \langle\mu_{\rm e}\rangle$ 	& $r_{\rm e}$ 			& $n$    			& $e_{\rm b}$ 			& $PA_{\rm b}$  
					&	$\mu_{0}$				& $h$    								& $e_{\rm d}$ 			& $PA_{\rm d}$   
					&	$\mu_{\rm bar}$   		& $a_{\rm bar}$  		& $n_{\rm bar}$  	& $e_{\rm bar}$ 	 	& $PA_{\rm bar}$  		& $c$      			& $B/T$  			& $Bar/T$		\\
$[$name]				&	[mag arcsec$^{-2}$]	& [mag arcsec$^{-2}$] 	& [arcsec]				&				&					& [$^{\circ}$]
					&	[mag arcsec$^{-2}$]		& [arcsec]								&					& [$^{\circ}$]
					&	[mag arcsec$^{-2}$]		& [arcsec]				&				&					& [$^{\circ}$]			&				&				&			\\
(1)	 				& 	(2)		 		& (3)		 			& (4)				& (5)		 			& (6) 
					& 	(7)					& (8)									& (9)					& (10)	
					& 	(11)					& (12)				& (13)			& (14)				& (15)				& (16)			& (17)			& (18)	& (19)	\\
\hline
SDSSJ1040 & ...  & ... & ...  & ...  & ...  & ...  & 17.611 $\pm$ 0.005 & 1.394 $\pm$ 0.008 & 0.238 $\pm$ 0.004 & 20.4 $\pm$ 0.5 & ...  &  ... &  ... & ...  & ...  & ...  & 0 & 0 \\ 
SDSSJ1121 & ... & ... &  ... & ...  &  ... &  ... & 18.299 $\pm$ 0.005 & 1.312 $\pm$ 0.007 & 0.050 $\pm$ 0.004 & 19.0 $\pm$ 0.5 & ...  &...   & ...  &  ... &  ... &...   & 0 & 0 \\ 
SDSSJ1136 &19.0 $\pm$ 0.1 & 18.3 $\pm$ 0.1 & 1.3 $\pm$ 0.1 & 1.06 $\pm$ 0.07 & 0.15 $\pm$ 0.02 & 172 $\pm$ 2 & 18.74 $\pm$ 0.04 & 3.17 $\pm$ 0.05 & 0.10 $\pm$ 0.01 & 128 $\pm$ 2 & ...  & ...  &  ... & ...  & ...  & ...  & 0.20 & 0 \\  
SDSSJ1152 & 20.1 $\pm$ 0.2 & 19.0 $\pm$ 0.5 & 2.9 $\pm$ 0.4 & 2.3 $\pm$ 0.2 & 0.75 $\pm$ 0.01 & 7 $\pm$ 2 & 20.77 $\pm$ 0.08 & 11.1 $\pm$ 0.3 & 0.23 $\pm$ 0.02 & 80 $\pm$ 2 & 20.87 $\pm$ 0.09 & 15 $\pm$ 2 & 6 $\pm$ 3 & 0.77 $\pm$ 0.02 & 32.6 $\pm$ 0.3 & 0.43 $\pm$ 0.06 & 0.10 & 0.03 \\ 
SDSSJ1332 & 20.2 $\pm$ 0.1 & 19.3 $\pm$ 0.2 & 1.3 $\pm$ 0.1 & 1.6 $\pm$ 0.1 & 0.050 $\pm$ 0.02 & 56 $\pm$ 2 & 19.12 $\pm$ 0.04 & 3.30 $\pm$ 0.06 & 0.038 $\pm$ 0.01 & 60 $\pm$ 2 & ...  & ...  &  ... & ...  & ...  & ...  & 0.12 & 0 \\ 
SDSSJ1432 & 19.2 $\pm$ 0.1 & 18.6 $\pm$ 0.1 & 0.85 $\pm$ 0.07 & 0.94 $\pm$ 0.06 & 0.38 $\pm$ 0.02 & 3 $\pm$ 2 & 18.53 $\pm$ 0.04 & 3.49 $\pm$ 0.06 & 0.080 $\pm$ 0.01 & 90 $\pm$ 2 & ...  &  ... & ...  &  ... & ...  & ...  & 0.04 & 0 \\ 
SDSSJ1444 & 18.7 $\pm$ 0.2 & 17.5 $\pm$ 0.5 & 0.63 $\pm$ 0.09 & 2.7 $\pm$ 0.3 & 0.15 $\pm$ 0.01 & 122 $\pm$ 2 & 20.18 $\pm$ 0.08 & 5.9 $\pm$ 0.1 & 0.21 $\pm$ 0.02 & 180 $\pm$ 2 & 21.22 $\pm$ 0.09 & 12 $\pm$ 2 & 2.0 $\pm$ 0.8 & 0.73 $\pm$ 0.02 & 159.3 $\pm$ 0.3 & 0.01 $\pm$ 0.01 & 0.12 & 0.06 \\ 
SDSSJ1603 & 19.2 $\pm$ 0.2 & 18.5 $\pm$ 0.2 & 0.92 $\pm$ 0.09 & 0.99 $\pm$ 0.07 & 0.186 $\pm$ 0.009 & 126 $\pm$ 1 & 19.82 $\pm$ 0.08 & 9.9 $\pm$ 0.2 & 0.061 $\pm$ 0.01 & 160 $\pm$ 1 & 21.99 $\pm$ 0.08 & 6 $\pm$ 1 & 0.11 $\pm$ 0.04 & 0.72 $\pm$ 0.01 & 102.0 $\pm$ 0.2 & 2.0 $\pm$ 0.2 & 0.03 & 0.01 \\ 
SDSSJ1701 & 17.5 $\pm$ 0.2 & 16.7 $\pm$ 0.4 & 0.54 $\pm$ 0.05 & 1.23 $\pm$ 0.09 & 0.217 $\pm$ 0.009 & 39 $\pm$ 1 & 19.87 $\pm$ 0.08 & 8.9 $\pm$ 0.2 & 0.11 $\pm$ 0.01 & 60 $\pm$ 1 & 21.03 $\pm$ 0.08 & 17 $\pm$ 3 & 2.0 $\pm$ 0.8 & 0.75 $\pm$ 0.01 & 69.3 $\pm$ 0.2 & $-1.0$ $\pm$ 0.1 & 0.06 & 0.03 \\ 
\hline  
\end{tabular}
\end{adjustbox}
\tablefoot{(1) Short name of the galaxy according to SDSS. (2), (3), (4),
(5), (6) and (7) Surface brightness at effective radius, mean surface 
brightness within effective radius, effective
radius, S\'ersic index, ellipticity ($e_{\rm bulge} = 1-q_{\rm
bulge}$) and position angle of the bulge, respectively. (8), (9),
(10) and (11) Central surface brightness, scalelength, ellipticity
($e_{\rm disk} = 1-q_{\rm disk}$) and position angle of the disk,
respectively. (12), (13), (14), (15), (16) and (17) Central surface
brightness, length, S\'ersic index, ellipticity ($e_{\rm bar} =
1-q_{\rm bar}$) , position angle and isophotal shape parameter of
the bar, respectively. (18) Bulge-to-total luminosity ratio. (19)
Bar-to-total luminosity ratio.} 
\label{tab:decomposition}
\end{sidewaystable}

\clearpage

\begin{figure*}[h!]
\centering
\includegraphics[width=17cm]{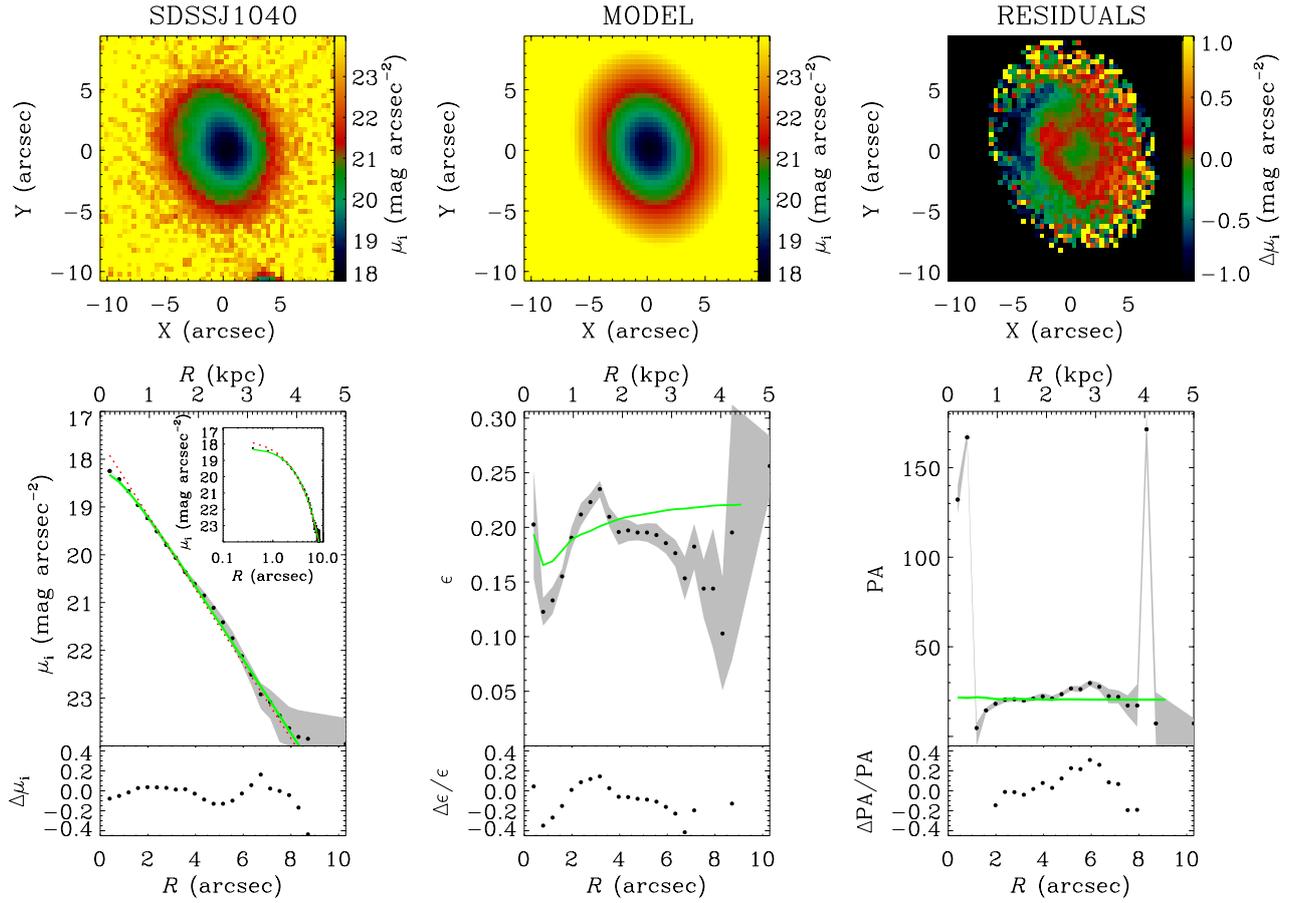}
\caption{Two-dimensional photometric decomposition of the $i$-band images 
of the sample galaxies obtained from GASP2D. For each galaxy the upper panels
(from left to right) show the map of the observed, modeled, and
residual (observed$-$modeled) surface brightness distributions. The
field of view is oriented with North up and East left. The black
areas in the residual image correspond to pixels excluded from the
fit. The lower panels (from left to right) show the ellipse-averaged
radial profile of surface brightness, position angle, and
ellipticity measured in the observed (black dots with grey error
bars) and seeing-convolved modeled image (green solid line) and their corresponding
difference. The intrinsic surface-brightness radial profiles of the
best-fitting bulge (blue dashed line), disk (red dotted line) and
bar component (magenta dotted-dashed line) are also shown in both
linear and logarithmic scale for the distance to the center of the galaxy.}
\label{fig:decomposition}
\end{figure*}

\begin{figure*}[t!]
\centering
\ContinuedFloat
\includegraphics[width=15cm]{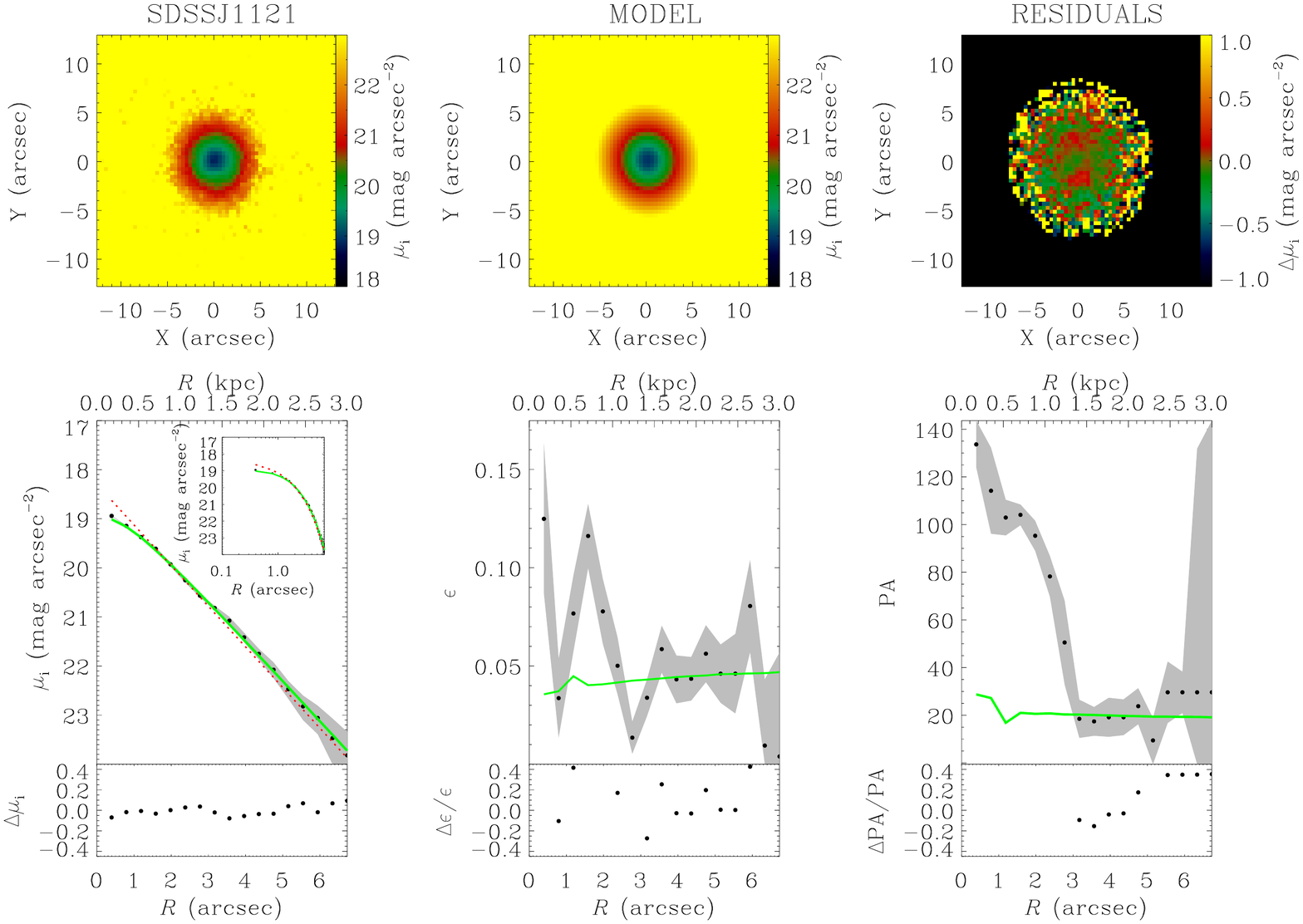}
\includegraphics[width=15cm]{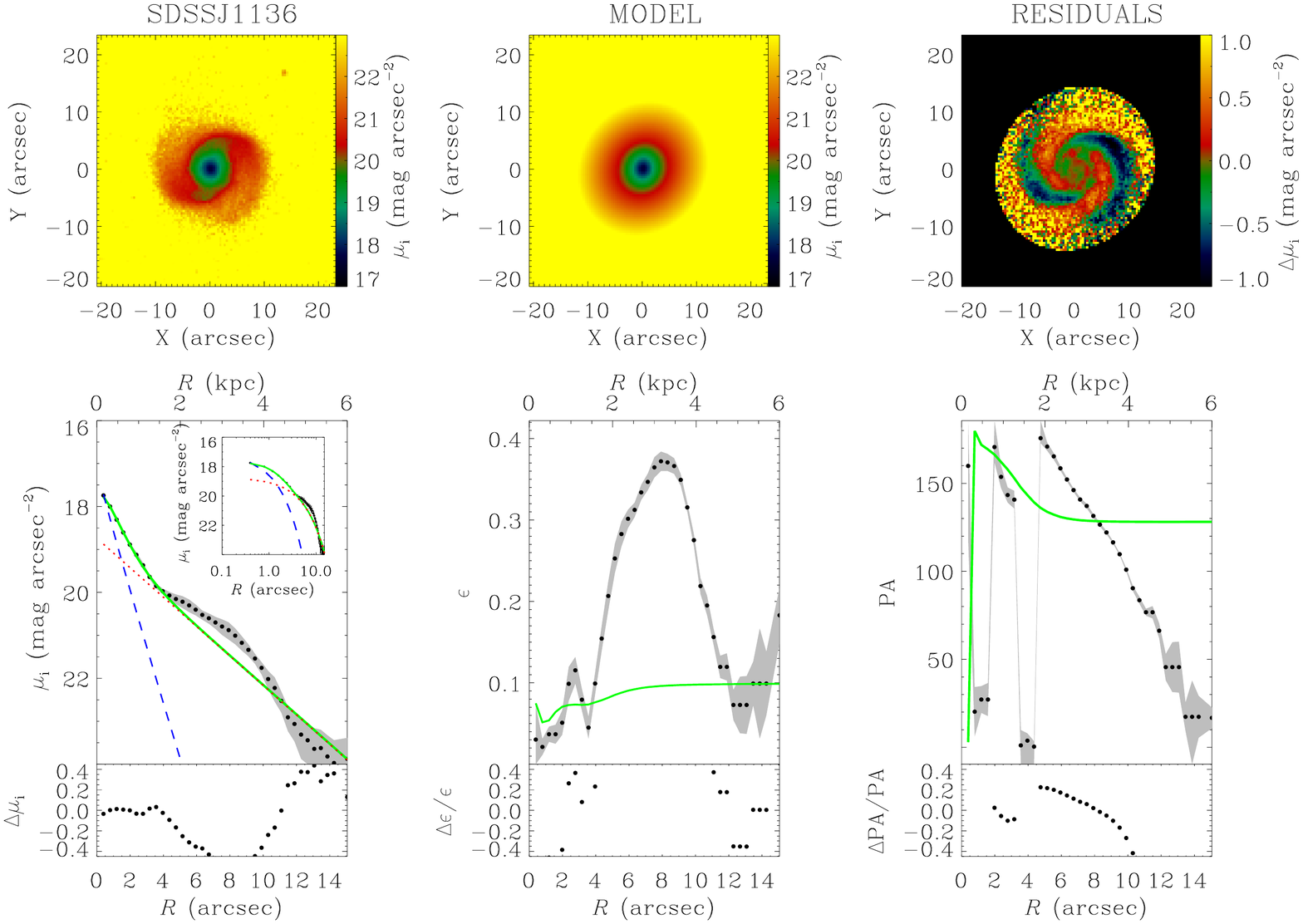}
\caption{continued.}
\end{figure*}

\begin{figure*}[t!]
\centering
\ContinuedFloat
\includegraphics[width=15cm]{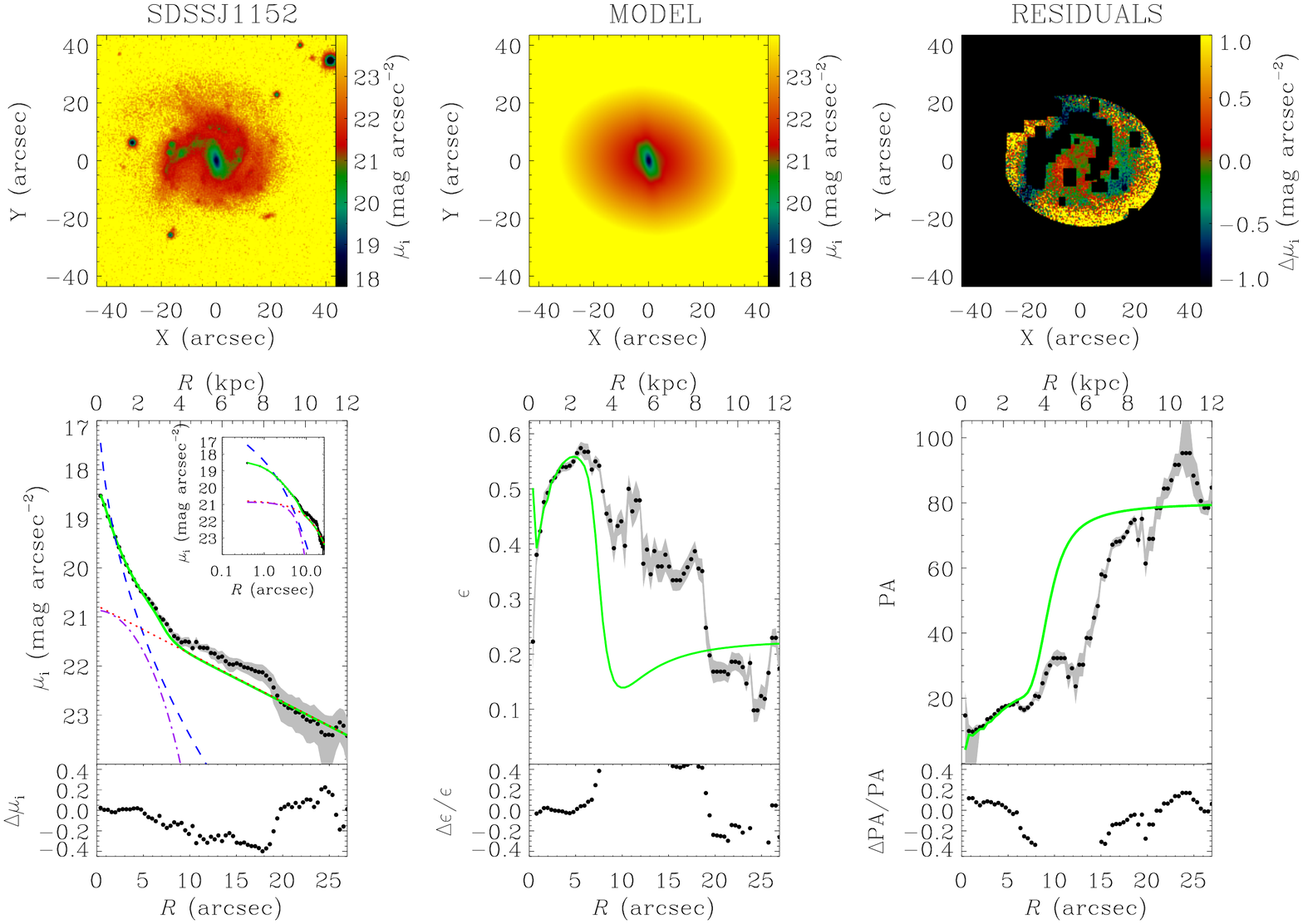}
\includegraphics[width=15cm]{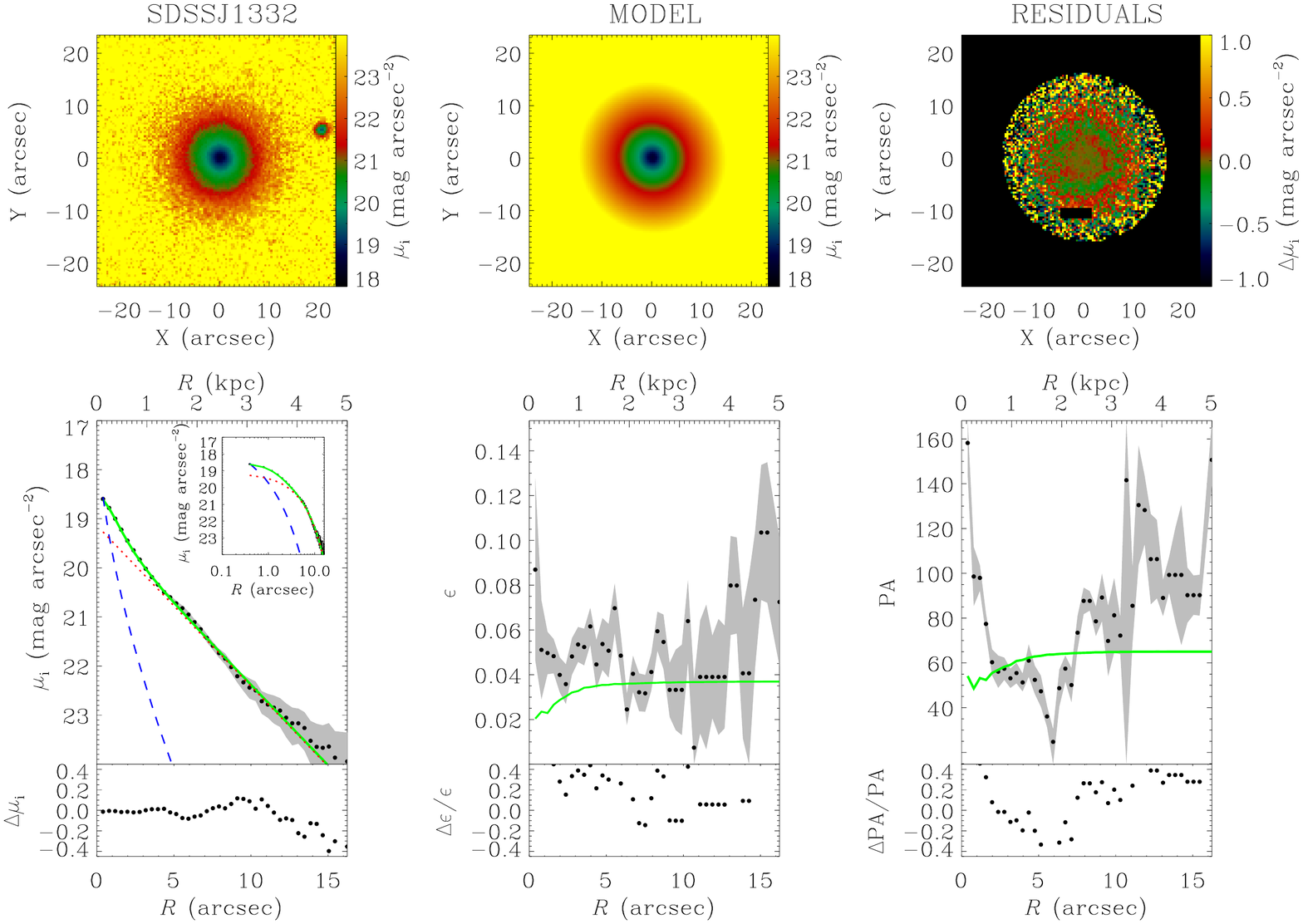}
\caption{continued.}
\end{figure*}

\begin{figure*}[t!]
\centering
\ContinuedFloat
\includegraphics[width=15cm]{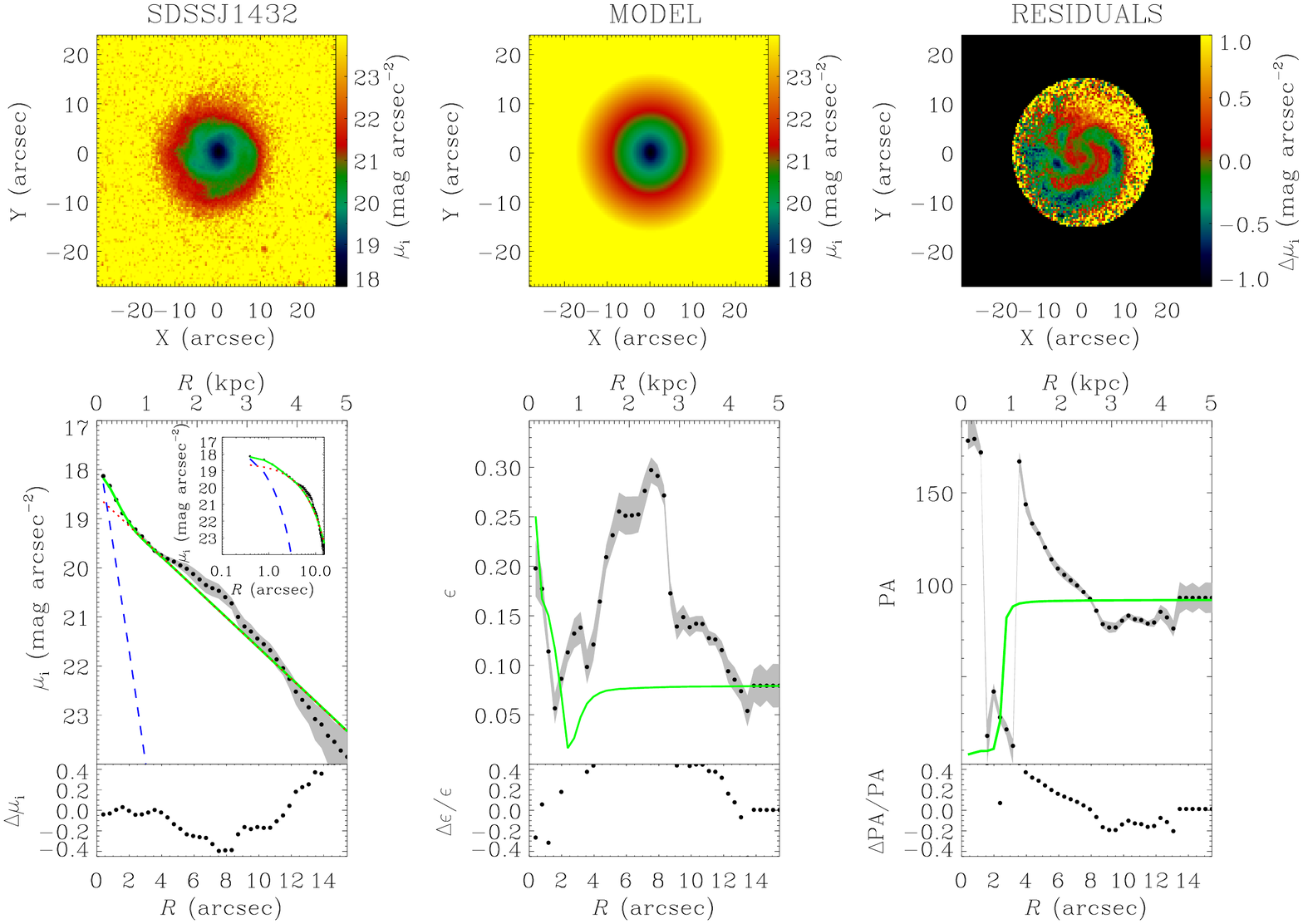}
\includegraphics[width=15cm]{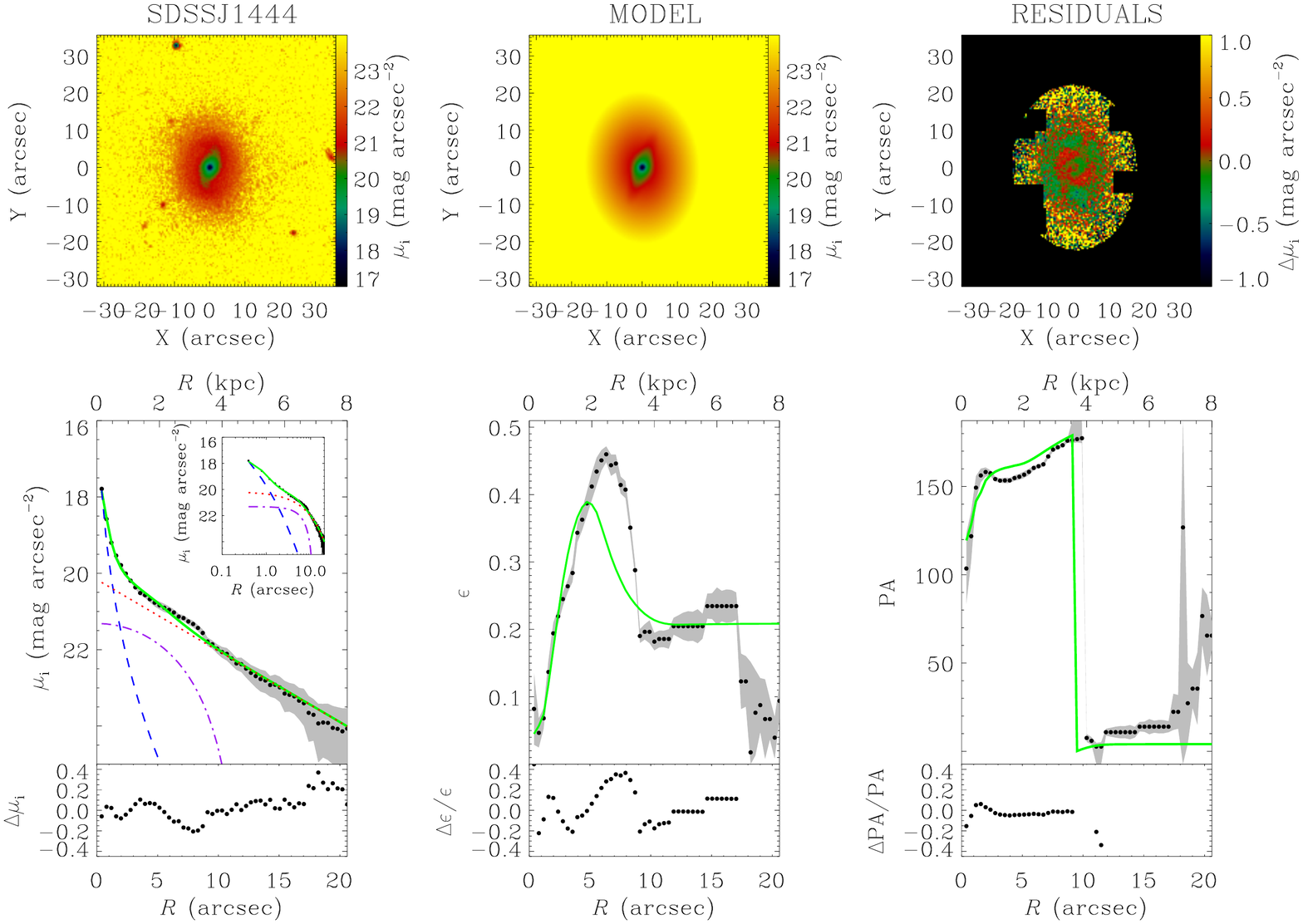}
\caption{continued.}
\end{figure*}

\begin{figure*}[t!]
\centering
\ContinuedFloat
\includegraphics[width=15cm]{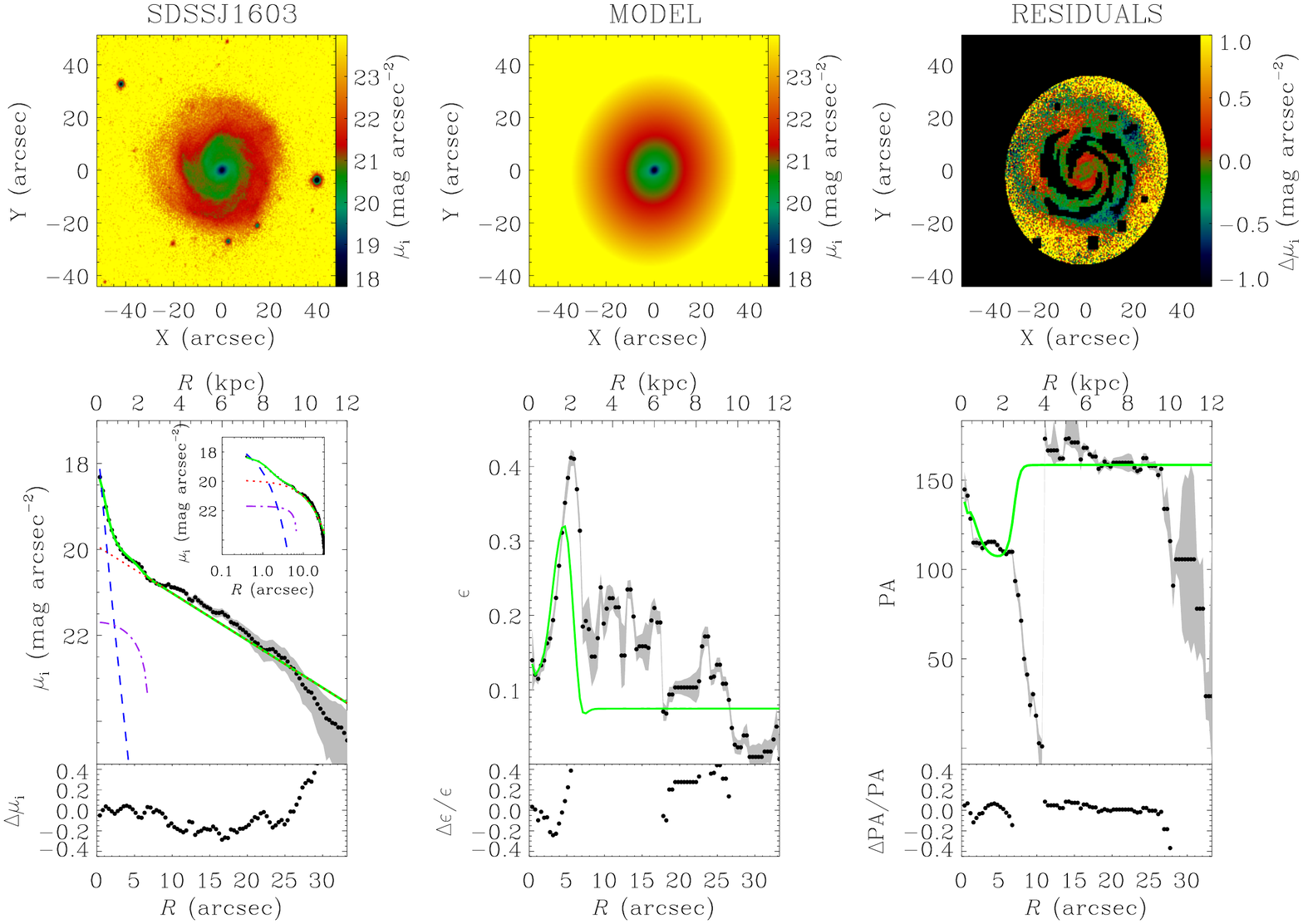}
\includegraphics[width=15cm]{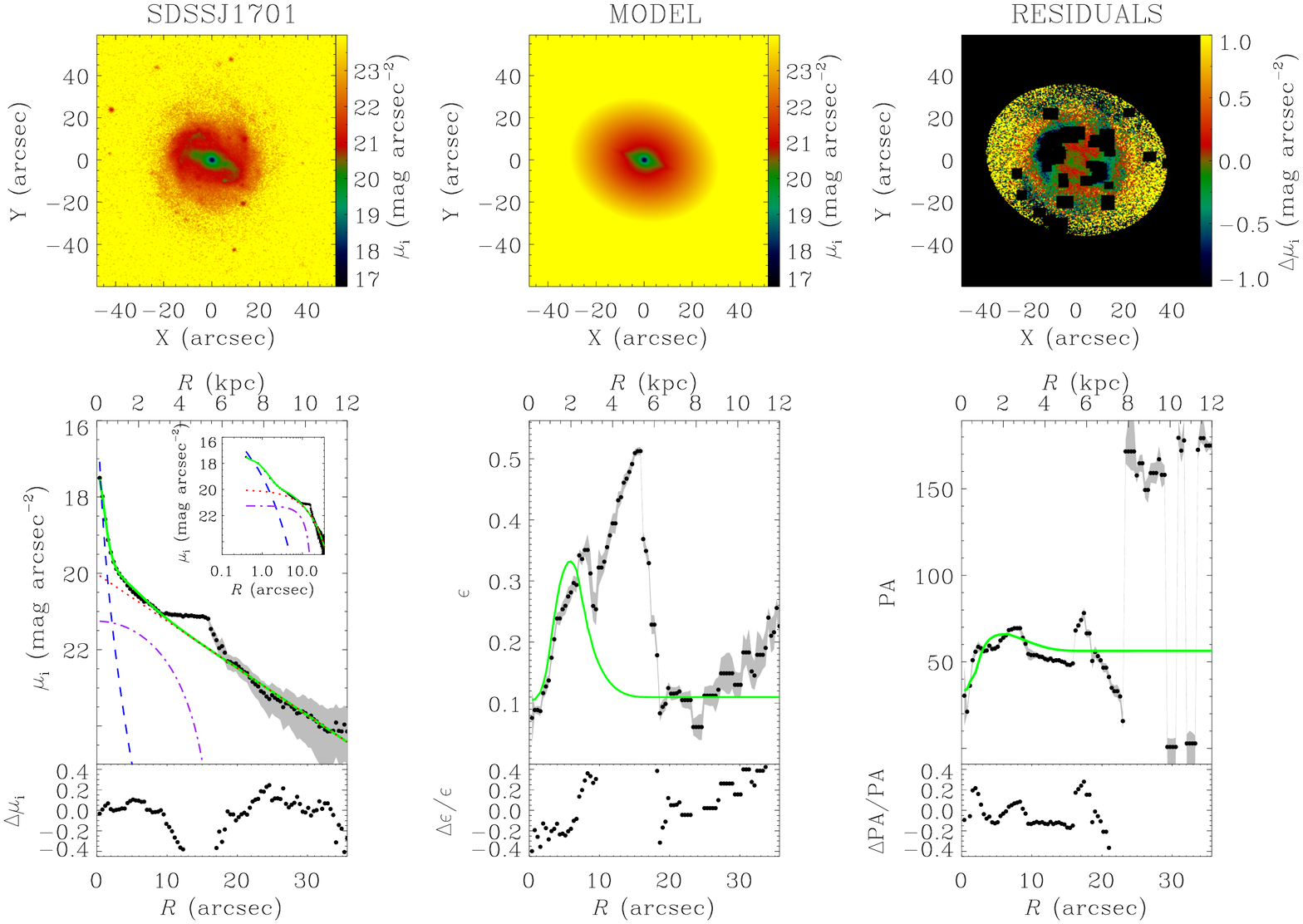}
\caption{continued.}
\end{figure*}

\clearpage


\section{Long-slit spectroscopy \label{sec:spectroscopy}}

\subsection{Observations and data reduction}

\begin{table*}[h!]
\centering
\caption{Log of spectroscopic observations and stellar kinematics of the sample galaxies}
\begin{tabular}{cccccc}
\hline
\hline
Galaxy & $PA$  & Single Exp. Time & Total Exp. Time & $\sigma_{\rm e}$ & $\sigma_{1.5 \, \rm arcsec}$\\
       & [$^{\circ}$]& [s]	       & [h] & [km~s$^{-1}$]	 & [km~s$^{-1}$]            \\
(1)    & (2)	    & (3)	       & (4) & (5) & (6)\\
\hline
SDSSJ1040 &     16.9 & $1\times1800 + 2\times2700$ & 2 & ...       & $54\pm4$  \\
SDSSJ1121 &     53.2 & $1\times1800 + 2\times2700$ & 2	& ...       & $37\pm4$  \\
SDSSJ1136 &     24.9 & $4\times1800              $ & 2	& $61\pm3$  & $61\pm3$   \\
SDSSJ1152 &     99.2 & $4\times1800    	  $ & 2	& $55\pm11$ & $58\pm11$  \\
SDSSJ1332 &     88.4 & $4\times1800    	  $ & 2 & $35\pm2$  & $35\pm2$ 	  \\
SDSSJ1432 &    101.8 & $1\times2400 + 1\times2700$ & 1.4 & $37\pm4$  & $42\pm3$ \\
SDSSJ1444 & $-$115.0 & $3\times1800 + 1\times2700$ & 2.25 & $55\pm9$  & $68\pm7$ \\
SDSSJ1603 &	  9.8 & $4\times1800    	  $ & 2 & $57\pm5$  & $56\pm4$	  \\
SDSSJ1701 &  $-$24.0 & $4\times1800              $ & 2 & $71\pm7$  & $69\pm6$ 	  \\
\hline  
\end{tabular}
\tablefoot{(1) Short name of the galaxy
according to SDSS. (2) Position angle of the slit measured North to
East. (3) Exposure time of the spectra. (4) Total exposure time. (5) Measured
velocity dispersion within $r_{\rm e}$. (6) Measured velocity
dispersion within 1.5 arcsec.}
\label{tab:log}
\end{table*}

We carried out the spectroscopic observations of the sample galaxies
on 2009 April 1-4 at the Telescopio Nazionale Galileo (TNG) in La
Palma (Spain).

In the setup we used the DOLORES spectrograph with the V510 grism
covering the wavelength range 4875-5325 \AA\ and the 1-arcsec wide
slit. The spectrograph was equipped with a E2V 4240 camera and a thinned
back-illuminated, deep-depleted, Astro-BB coated CCD with
$2048\times2048$ pixels of $13.5\times13.5$ $\mu$m$^2$, gain 0.97
$e^{-}$~ADU$^{-1}$, and readout noise 9 $e^{-}$ (rms).  We used the
mean of the Gaussian FWHMs measured for a number of unblended arc-lamp
lines over the whole spectra range of a wavelength-calibrated spectrum
to derive the instrumental resolution. We found $\rm FWHM_{\rm inst} =
1.04\pm0.08$ \AA, corresponding to a velocity dispersion $\sigma_{\rm inst} = 25.4 \pm 0.4$ km~s$^{-1}$ at 5100 \AA, far below the
instrumental resolution of the SDSS spectra. The angular sampling was
0.252 arcsec~pixel$^{-1}$, with a reciprocal dispersion of 0.235 \AA~pixel$^{-1}$.  
The median value of the seeing FWHM during
the observing nights was 1.22 arcsec. It was measured fitting a
circular Gaussian on the guide stars.

For each object, we centered the slit on the galaxy nucleus 
and visually aligned it along the galaxy major axis. Details on the slit position
angles and exposure times are given in Table
\ref{tab:log}.

Each spectrum was bias-subtracted, flat-field corrected, cleaned of
cosmic rays, corrected for bad columns, and wavelength and flux
calibrated using standard IRAF tasks. We used the bias frames
obtained during the observing nights to determined the bias level. We
performed the flat-field correction for pixel-to-pixel sensitivity
variations and large-scale illumination patterns arising from slit
vignetting using thorium lamp and twilight sky spectra respectively, which
were normalized and divided into all spectra.  We carried out the
identification of the cosmic rays by comparing the counts in each
pixel with the local mean and standard deviation as obtained from
Poisson statistics considering the gain and readout noise of the CCD
and then we corrected by interpolating over. If residual cosmic rays
were present, we manually removed them by editing the spectra.  We
rebinned all the spectra using the wavelength solution obtained from
the corresponding arc-lamp spectrum and flux-calibrated using the
sensitivity function acquired from the flux standard star spectra of
the corresponding night. We estimated
the contribution from the sky by interpolating along the outermost
regions at the two edges of the slit, where the galaxy or stellar
light is negligible, and then we subtracted it. 
Finally, in order to improve the $S/N$ of the final
two-dimensional spectrum, we coadded the major-axis spectra using the
center of the stellar continuum as reference. 

\subsection{Stellar kinematics}

\begin{figure*}[t!]
\centering
\includegraphics[width=17cm]{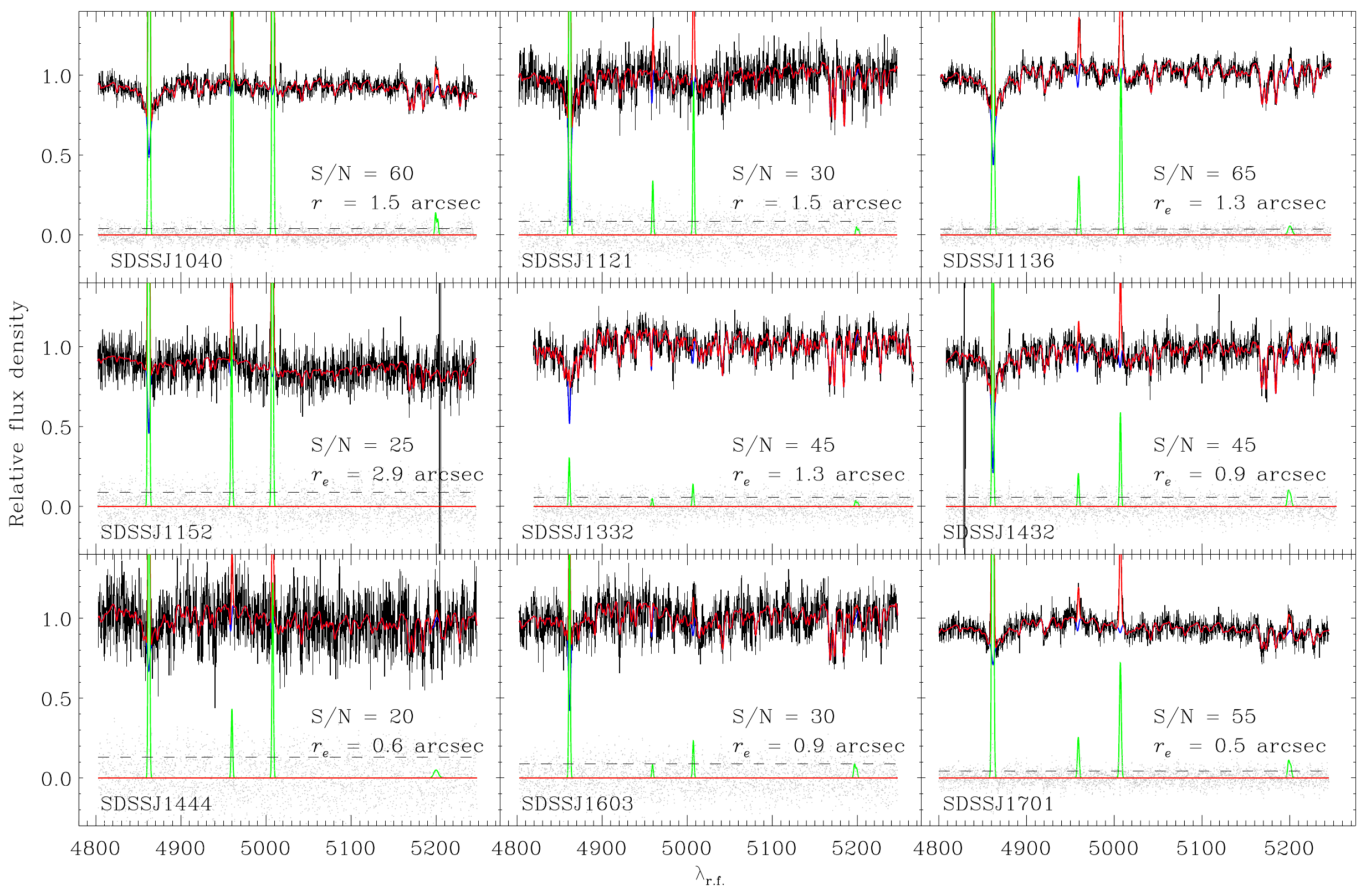}
\caption{Spectra of the sample galaxies. 
Relative fluxes have false zero points for viewing convenience. 
In each panel the best-fitting model (red line) is the sum of 
the spectra of the ionized-gas (green line) and stellar 
component (blue line).
The residuals (grey dots) are obtained by subtracting the best-fitting
model from the observed spectrum.
The $S/N$ is given per resolution element.}
\label{fig:spectra}
\end{figure*}

We measured the stellar kinematics of the sample galaxies from the
absorption features in their spectra.  We used the penalized
pixel-fitting method \citep[pPXF;][]{Cappellari2004}, including the
Gas and Absorption Line Fitting algorithm
\citep[GANDALF;][]{Sarzi2006}, adapted for dealing with our setup.  We
rebinned the spectra along the spatial direction in order to provide
the kinematic parameters within 1.5 arcsec and $r_{\rm e}$.

The convolution of a linear combination of stellar spectra from the
ELODIE library \citep[FWHM = 0.48 \AA;][]{Prugniel2001} with the
line-of-sight velocity distribution (LOSVD), described by
Gauss-Hermite expansion \citep{Gerhard1993, vanderMarel1993}, allowed us to
fit the observed galaxy spectrum by $\chi^2$ minimization in pixel
space.  Before the convolution, we degraded the spectral resolution of
the ELODIE spectra by convolving them with a Gaussian function in
order to match the galaxies spectral resolution.  We properly masked
bad pixels coming from imperfect subtraction of cosmic rays and sky
emission lines and we excluded them from fitting procedure.  In this
way, we determined the value of the mean velocity $v_{\rm los}$,
velocity dispersion $\sigma_{\rm los}$, and third- $h_3$ and
fourth-order $h_4$ moments of the LOSVD.  The measured values of $h_3$ and
$h_4$ were compatible with zero, thus we fixed their values and we
performed again the fit adopting a Gaussian LOSVD and obtaining only
$v_{\rm los}$ and $\sigma_{\rm los}$. 
All the analyzed spectra had a $S/N \ge 20$ per resolution element (Fig. \ref{fig:spectra}).
We adopted a low-order multiplicative polynomial in the template fitting, 
in order to deal with the effect of dust and possible residuals of the 
data reduction procedure. Thus, the method minimizes the effects of 
reddening, being more sensitive to the absorption lines than the continuum shape.

To allow for additional structure that is not addressed by our
model, we estimated the uncertainties in the kinematic parameters from
the formal errors of the fitting procedure by evaluating the $\chi^2$ values. 
We achieved $\chi^2 = N_{\rm dof} = N_{\rm data} - N_{\rm fit}$ for the best-fitting model,
where $N_{\rm dof}$, $N_{\rm data}$, and $N_{\rm fit}$ are the
numbers of the degrees of freedom, data points, and fitting
parameters, respectively \citep{Press1992}.
The measured stellar velocity dispersions and
corresponding errors are reported in Table \ref{tab:log}.


\section{Scaling relations \label{sec:scaling_relations}}

The luminosity, surface brightness, size, and velocity dispersion of
ellipticals and bulges of disk galaxies are used to identify a set of
well-defined scaling relations, namely the FPR, KR, and FJR. These
relations unveil the structure of galaxy spheroids and provide
valuable clues on the physics driving their formation and evolution
history.

We investigated whether small bulges follow the same scaling relations
traced by ellipticals and large bulges by comparing our galaxy sample
to the sample of nearby galaxies with structural parameters and
velocity dispersion measured by \citet{Gadotti2009} and \citet{Oh2011} from SDSS \emph{i}-band
images and spectra, respectively.

We marked all the comparison galaxies
 by assigning a probability to be E-S0, Sa-Sb or Sc-Sd using the
Bayesian automated classification by \citet{HuertasCompany2011}.
After that, ellipticals ($B/T$ = 1) were disentangled from lenticulars ($B/T$ < 1). 
This allowed us to properly consider the structural parameters of the whole
galaxy for ellipticals and of the bulge component only for disk galaxies.

We are particularly interested in having low-$\sigma$ galaxies in the
comparison sample. But, given the typical $S/N$ and instrumental
resolution of the SDSS spectra ($\sigma_{\rm inst, SDSS} \simeq 70$ km
s$^{-1}$), as a rule SDSS galaxies with $\sigma<70$ km
s$^{-1}$ are recommended to not be used because their velocity
dispersion could be unreliable \citep{Bernardi2003a}. \citet{Oh2011}
defined a new criterion to assess the reliability of the SDSS-based
$\sigma$ values by rejecting galaxies with a ratio rN/sN $> 3$ between the rms
of the residuals of the spectral fit (rN) and the expected
statistical rms (sN). We eventually included in the
comparison sample all the galaxies with $\sigma< 70$ km s$^{-1}$ which
meet the prescription by \citet{Oh2011}.

For all the comparison galaxies we calculated the velocity dispersion
$\sigma_{\rm e}$ within $r_{\rm e}$ from the SSDS $\sigma$ measured
within a fixed aperture of 3 arcsec using a power-law function
\begin{equation}
\left(\dfrac{\sigma}{\sigma_{\rm e}} \right) = 
  \left(\dfrac{r}{r_{\rm e}} \right)^{\alpha} \, ,
\end{equation}
where $\alpha$ is
\begin{equation*}
\alpha_{\rm ETG} = -0.055 \pm 0.020
\end{equation*}
for the early-type galaxies (ETGs), and 
\begin{equation*}
\begin{cases}
\alpha_{\rm LTG} (M_r < -22) = 0.047 \pm 0.021 \\
\alpha_{\rm LTG} (-20 < M_r < -22) = 0.086 \pm 0.013 \, \, , \\ 
\alpha_{\rm LTG} (M_r > -20) = 0.153 \pm 0.063
\end{cases}
\end{equation*}
for the late-type galaxies (LTGs), as derived by
\citet{FalconBarroso2016} from the analysis of 300 galaxies drawn from
the Calar Alto Legacy Integral Field Area Survey (CALIFA) DR3
\citep{Sanchez2016}.

\begin{figure*}[t!]
\centering
\includegraphics[width=17cm]{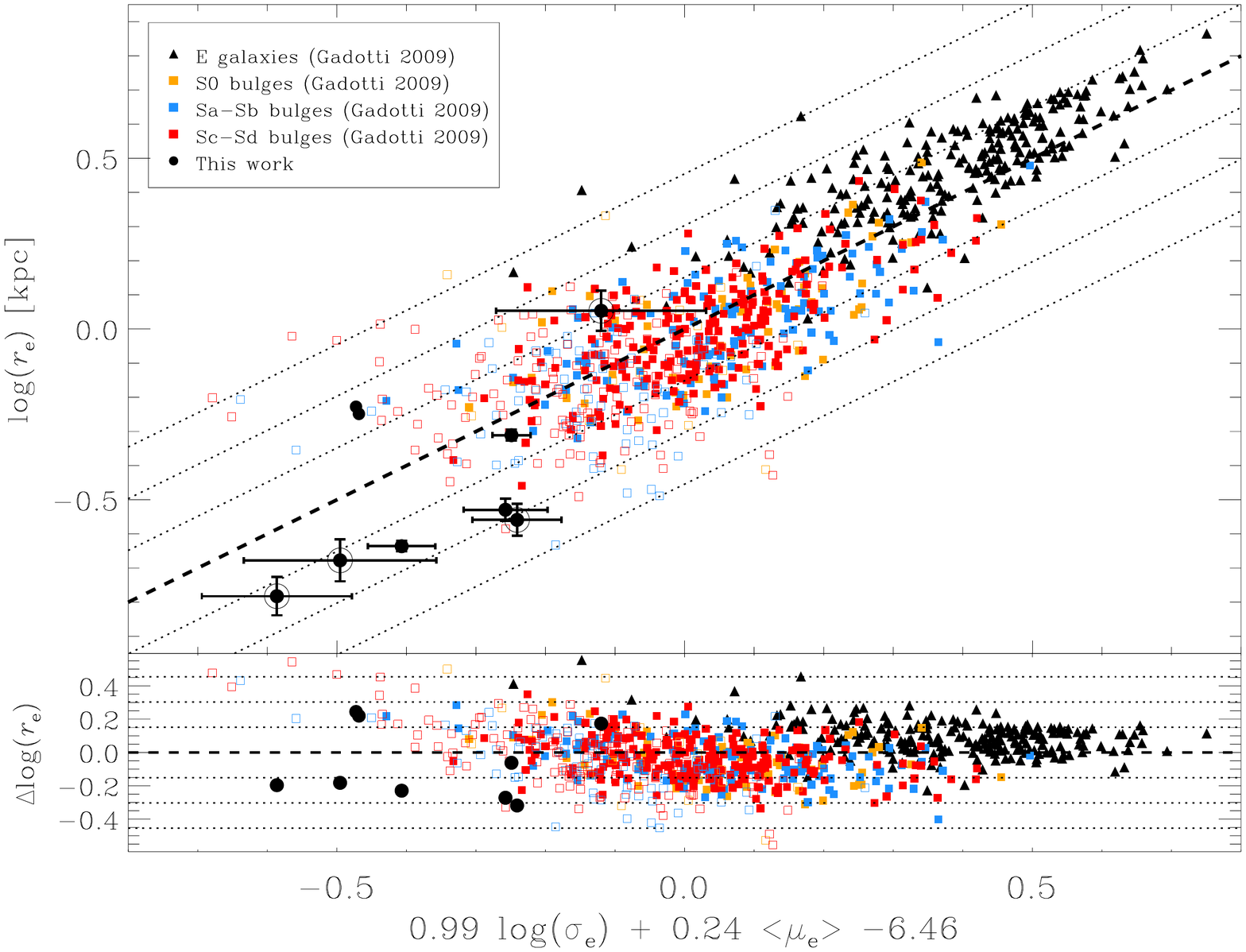}
\caption{FPR for the galaxies of our (larger symbols) and comparison
sample (smaller symbols). The ellipticals (black triangles) and
bulges in S0 (orange squares), Sa-Sb (blue squares), and Sc-Sd
galaxies (red squares) of the comparison sample are shown with
filled ($\sigma>70$ km s$^{-1}$) and empty symbols ($\sigma<70$ km
s$^{-1}$) according to their calculated velocity dispersion. Filed circles
with and without error bars correspond to the galaxies in our sample
with and without bulge, respectively. Open circles mark the barred
galaxies. The dashed line is the
best-fitting relation for the comparison sample. The dotted lines
show the $1$ rms, $2$ rms, and $3$ rms deviation in
$\log{(r_{\rm e})}$ regions, respectively.}
\label{fig:fundamental_plane}
\end{figure*}

At the end, the comparison sample includes 234 ellipticals and 626
bulges (79 in lenticulars, 192 in Sa-Sb galaxies, 355 in Sc-Sd
galaxies).
For the sake of completeness, although only 7
out of 9 galaxies in our sample present a bulge component, all of
them were superimposed on the scaling relations for a comparison.
Due to the small size of our sample, in the following we discuss only
the properties of our individual late-type bulges and not of the whole
population. Further observations of a complete sample of late-type
bulges would be needed to infer their global properties as a distinct
class of bulges.

\subsection{Fundamental plane \label{subsec:fundamental_plane}}

\begin{figure*}[t]
\centering 
\includegraphics[width=17cm]{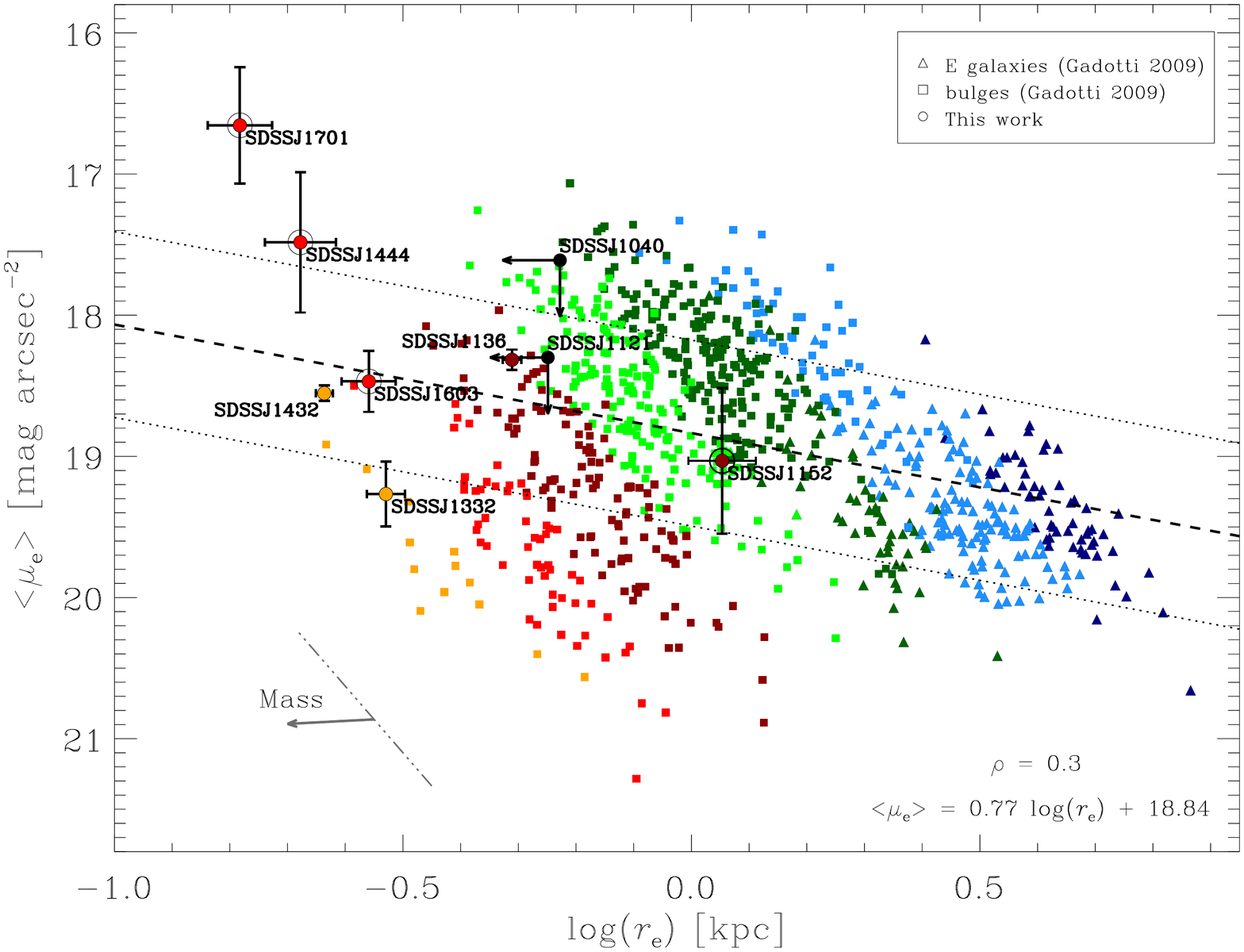}
\caption{KR for the galaxies of our (larger symbols) and comparison
sample (smaller symbols). The ellipticals and bulges of the
comparison sample are shown with triangles and squares,
respectively. Filled circles with and without error bars correspond to the
galaxies in our sample with and without bulge, respectively. 
Open circles mark the barred galaxies. The galaxies are
divided according to their absolute magnitude in the following bins:
$M_i< -22$ mag (dark blue), $M_i = [-22, -21]$ mag (light blue),
$M_i = [-21, -20]$ mag (dark green), $M_i = [-20, -19]$ mag (light
green), $M_i = [-19, -18]$ mag (dark red), $M_i = [-18, -17]$ mag
(light red), and $M_i = [-17, -16]$ mag (orange).  The dashed line
is the best-fitting relation for the comparison sample. The dotted
lines show the rms deviation in $\langle \mu_{\rm e} \rangle$ from
the fit. The dash-dotted line gives the slope of the KR for the
magnitude bin $M_i = [-20, -19]$ mag, while the arrow indicates the
KR trend for decreasing masses \citep{NigocheNetro2008}.  
The best-fitting relation and Pearson correlation
coefficient are also given.}
\label{fig:kormendy}
\end{figure*}

The FPR \citep{Djorgovski1987, Dressler1987, DOnofrio2008, Cappellari2013} is the
most widely studied scaling relation providing information about
structure and kinematics of galaxy spheroids.  Observable quantities,
such as the mean effective surface brightness $\langle \mu_{\rm e}
\rangle$, $r_{\rm e}$ and $\sigma$, are commonly adopted in the FPR as proxies
of the physical properties of the galaxy 
\citep{Bender1992}. When spirals are considered, as in this work,
the previous properties refer to the bulge component only instead of to the
whole galaxy like in ellipticals.

The common way to express the FPR is $\log(r_{\rm e}) = \alpha
\log(\sigma_{\rm e}) + \beta \langle \mu_{\rm e} \rangle + \, \gamma
\, $ in order to separate $r_{\rm e}$ from $\sigma_{\rm e}$ and
$\langle \mu_{\rm e} \rangle$ which do not depend on distance.
Fig. \ref{fig:fundamental_plane} shows the FPR for the comparison sample of
ellipticals and bulges with the best-fitting line given by
\begin{equation}
\log(r_{\rm e}) = 0.99 \log(\sigma_{\rm e}) + 0.24  \langle \mu_{\rm e} \rangle - \, 6.46 \, .
\end{equation}
where $r_{\rm e}$ is given in kpc, $\sigma_{\rm e}$ in km s$^{-1}$,
and $\langle \mu_{\rm e} \rangle$ in mag
arcsec$^{-2}$. We derived the FPR coefficients with a direct fit
approach using the \texttt{cramer} routine in IDL. 
They are consistent with those found by \citet{Bernardi2003b}
considering a sample of 8022 ETGs and analyzing their \emph{i}-band
images from the SDSS. The rms deviation in $\log{(r_{\rm e})}$ from
the fitted relation ($\rm rms=0.15$) is larger than the one obtained
by \citet{Bernardi2003b} but similar to the \emph{K}-band and
\emph{B}-band values of \citet{FalconBarroso2002} and to the
\emph{V}-band value of \citet{FalconBarroso2011}, respectively. 
In literature there is a general agreement that the typical scatter of the FPR ($\rm
rms\simeq0.10$) is independent of the photometric passband
\citep{Jorgensen1996, LaBarbera2010, Kim2016}. But, it should be
remarked that most works concentrate on ellipticals and lenticulars
and adopt different fitting methods.

As a matter of fact, it is observationally easier to study the large and bright bulges of
lenticulars and early-type spirals than the small and faint bulges of
late-type spirals, which are more affected by dust patches and lanes and
require more complex photometric decompositions
\citep{Laurikainen2010}. Since they share similar properties, ellipticals and
early-type bulges are usually considered to follow the same FPR in
optical and near-infrared passbands, whereas late-type bulges deviate from the relation \citep{FalconBarroso2002}.
Some authors explained the deviation of late-type bulges from the FPR
by taking into account the total kinetic energy of the system
\citep{Bender1992, FalconBarroso2002}. Correcting the velocity dispersion for this missing
contribution allows to place bulges of different morphological types
on the same FPR \citep{Bender1992, FalconBarroso2002}.
We measured the rotational velocity of
our sample bulges within $r_{\rm e}$. After correcting for slit orientation and galaxy
inclination, we found that the maximum rotational velocity $v_{\rm
max}$ is much smaller than the corresponding
$\sigma_{\rm e}$, being $v_{\rm max}\simeq50$ km s$^{-1}$ for SDSSJ1152 and $|v_{\rm max}|<30$ km s$^{-1}$ for the
remaining galaxies. 
The slit misalignment does not affect $\sigma_{\rm e}$ 
which we measured over nearly squared apertures.
Thus, the correction of the velocity dispersion values
did not affect the position of the bulges in the FPR, although these measurements are
subject to our observational limits in terms of both the large PSF
FWHM of our spectra with respect to $r_{\rm e}$ and possible
contamination from the underlying disk.
However, the surface-brightness distribution is dominated by the
bulge contribution by a factor 2-10 with respect to the disk inside
$r_{\rm e}$. Therefore, we are confident that our
luminosity-weighted spectroscopic measurements are probing the bulge
kinematics.

Each of our bulges is consistent with the FPR of the comparison sample
(Fig. \ref{fig:fundamental_plane}).
No obvious differences appear in the residuals when
galaxies are divided according to their morphological type.
Moreover, the FPR suggests that a single population of 
bulges share the same physical properties. 
This is also true when our small bulges are considered.

\subsection{Kormendy relation}

\begin{figure*}[t!]
\centering
\includegraphics[width=17cm]{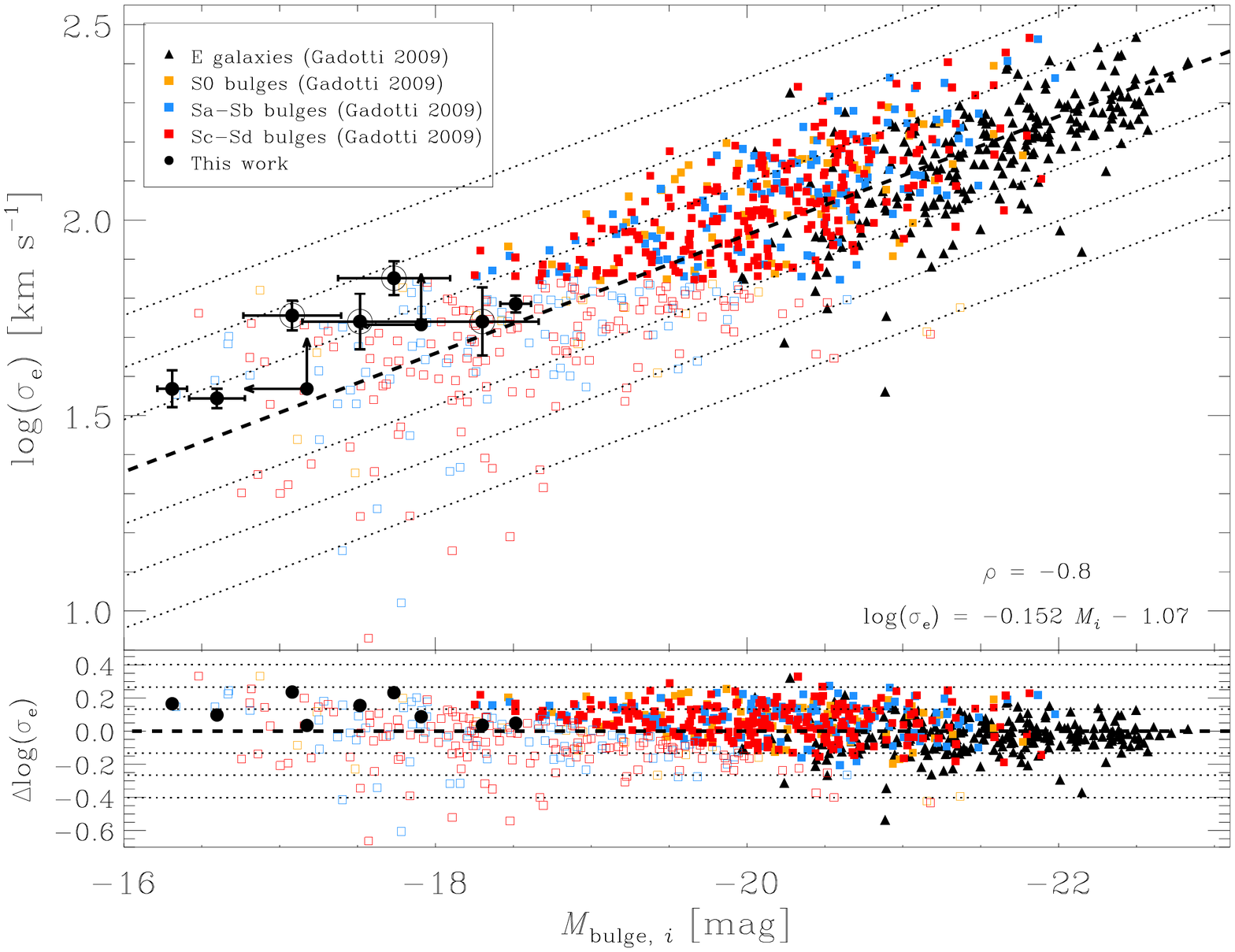}
\caption{As in Fig.~\ref{fig:fundamental_plane} but for the FJR and
with the rms deviation in $\log{(\sigma_{\rm e})}$ from the fit. The
best-fitting relation and Pearson correlation coefficient are also
given.}
\label{fig:faber_jackson}
\end{figure*}

According to the KR \citep{Kormendy1985, NigocheNetro2008} larger galaxies are fainter
than smaller ones. Fig. \ref{fig:kormendy} shows the KR for the
comparison sample of ellipticals and bulges with the
best-fitting line given by
\begin{equation}
\langle \mu_{\rm e} \rangle \, = 0.77 (\pm 0.08) \log(r_{\rm e}) + 18.84(\pm 0.02)  \, ,
\end{equation} 
where $\langle \mu_{\rm e} \rangle$ is expressed in mag arcsec$^{-2}$
and $r_{\rm e}$ in kpc. 
We derived the KR coefficients using the \texttt{poly\_fit} routine in IDL.
The low value of the Pearson correlation
coefficient ($\rho = 0.3$) reflects the large scatter of the relation
when all the comparison galaxies are considered together. The rms
deviation in $\langle \mu_{\rm e} \rangle$ from the fitted relation
($\rm rms=0.7$ mag arcsec$^{-2}$) is slightly larger than values
quoted in earlier works \citep[e.g.,][]{Hamabe1987, LaBarbera2003}
making it difficult to observe a clear trend, contrary to previous
findings \citep[e.g.,][]{Aguerri2004, Ravikumar2006}. 
However, comparing galaxies with different absolute
magnitudes could be misleading because of the strong bias caused by
selecting systems with different stellar masses
\citep{NigocheNetro2008}. Indeed, we found a clear trend when
different intervals of absolute magnitude are considered in agreement
with \citet{NigocheNetro2007, NigocheNetro2008}, who pointed out how
the KR coefficients and rms deviation from the fitted relation change
when choosing fixed-width intervals of progressively brighter absolute
magnitude.
  
Our bulges are consistent with the magnitude-dependent trend of the KR
and 5 of them lie in the poorly populated region of the low-mass
systems characterized by small $r_{\rm e}$ and large $\langle \mu_{\rm
e} \rangle$ (Fig. \ref{fig:kormendy}). 
The offset of the bulge of SDSSJ1152 is explained by the large
errors on $r_{\rm e}$ and $\langle \mu_{\rm e} \rangle$. Indeed, the
data point is consistent within $3\sigma$ with its magnitude bin.
Due to the large scatter and
overall dependance on the mass of the KR, we can not infer any difference
in the populations of bulges using such a relation, which is indeed a poor 
proxy to disentangle bulge properties.

\subsection{Faber-Jackson relation}

\begin{figure*}[t!]
\centering
\includegraphics[width=17cm]{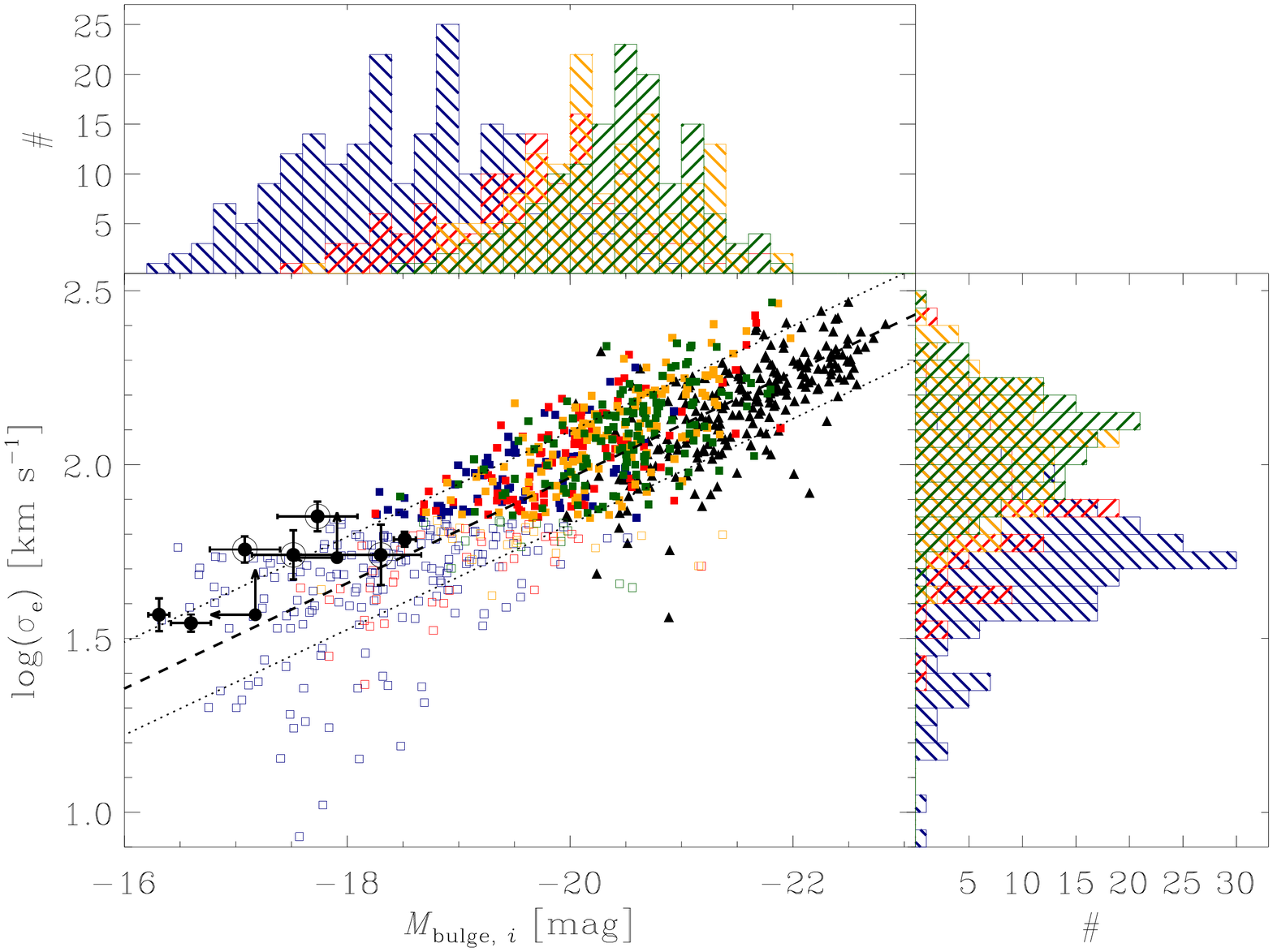}
\caption{FJR for the galaxies of our (larger symbols) and comparison
sample (smaller symbols). The ellipticals (triangles) and bulges
(squares) of the comparison sample are shown with filled
($\sigma>70$ km s$^{-1}$) and empty symbols ($\sigma<70$ km
s$^{-1}$) according to their calculated velocity dispersion.
Filled circles with and without error bars correspond to the galaxies in our
sample with and without bulge, respectively. 
Open circles mark the barred galaxies. The galaxies are
divided according to their S\'ersic index in the following bins:
$1<n<2$ (blue), $2<n<3$ (red), $3<n<4$ (orange), and $n>4$ (green).
The dashed line is the best-fitting relation for the comparison
sample. The dotted lines show the rms deviation in
$\log{(\sigma_{\rm e})}$ given in Fig.~\ref{fig:faber_jackson}.  The
distribution of galaxies of the different bins of $n$ as function of
their absolute magnitude and effective velocity dispersion are also
shown.}
\label{fig:faber_jackson_n}
\end{figure*}

The FJR \citep{Faber1976, FalconBarroso2011} states that brighter galaxies
exhibit larger velocity dispersion.

Fig. \ref{fig:faber_jackson} shows the FJR for the comparison sample
of ellipticals and bulges with the best-fitting line given by
\begin{equation}
\log(\sigma_{\rm e}) = -0.152 (\pm 0.003) M_{i} - 1.07 (\pm 0.07) \, ,
\label{eq:faber_jackson}
\end{equation}
where $\sigma_{\rm e}$ is provided in km s$^{-1}$ and $M_i$ is the
\emph{i}-band absolute magnitude of the galaxy spheroid.  
We derived the FJR coefficients using the \texttt{poly\_fit} routine in IDL.
 From Eq.~\ref{eq:faber_jackson} it results
$L\propto\sigma^{2.64 \pm 0.01}$, which deviates from the theoretical
virial relation $L\propto\sigma^{4}$ but it is consistent with
$L\propto\sigma^{2.9 \pm 0.5}$ found by \citet{Balcells2007} for a
sample of bulges observed with the Hubble Space Telescope in the
\emph{K} band. The Pearson correlation coefficient is $\rho = -0.8$,
reflecting a tight correlation between galaxies properties in this plane.  

The scatter increases in the low-$\sigma$ end of the relation, showing a trend 
mimicking the expected down-bending of the FJR \citep{FalconBarroso2011}. 
This regime of the FJR is populated by the small bulges of the
comparison sample. Their velocity dispersions are far below the
instrumental resolution of the SDSS spectra.
Moreover, they are measured within a fixed aperture of 3
arcsec, where a significant contamination from the disk component is
expected due to the small $B/T$ of these galaxies. On the contrary, no
hint of down-bending is observed for our bulges, which have measured
$\sigma$ larger than instrumental resolution. At face
values, they are above the FJR whereas disk-like bulges are expected to be
low-$\sigma$ outliers, being rotation rather than pressure supported.
As a matter of fact, each of our bulges
follows the same FJR of the brighter ellipticals and bulges of the
comparison sample. Moreover, we can infer that both ellipticals and bulges 
share the same properties when the FJR is used.
The FJR down-bending highlighted in other works
\citep{Mendez2008a, Kim2016} could be due to
selection effects, when only high- or low-mass systems are considered.
Indeed, the slope of the relation changes when galaxies with
increasing values of absolute magnitudes are considered
\citep{NigocheNetro2010}.

We did not observe any trend when bulges are divided according to the
morphological type of their host galaxies (Fig
\ref{fig:faber_jackson}) in agreement with the results by
\citet{Thomas2006}.
They found that the stellar populations of less massive bulges are
typically younger, less metal-rich, and less overabundant of $\alpha$
elements than ETGs with no dependance on the morphological type. This
also supports the correlation between the structural parameters (e.g.,
$r_{\rm e}/h$ and $B/T$) and bulge luminosity and mass rather than
morphological type.

We observed a smooth transition from larger to smaller values of $n$
according to both $\sigma$ and $M_{{\rm bulge},\, i}$ when the sample is divided in
terms of the S\'ersic index (Fig. \ref{fig:faber_jackson_n}). This
suggests that $n$ strongly depends on the mass of the system.  The
same trend is visible when $B/T$ is considered (see Equation \ref{eq:n_BT} in Sect. \ref{sec:err_bud}).
Recently, \citet{Kim2016} have shown that 
faint galaxies ($-19 < M_{r, \, \rm bulge} < -17$ mag) with smaller $B/T$ have
systematically lower $\sigma$ than those with larger $B/T$. 
They concluded that bulges in low-$B/T$ galaxies are rotation supported (i.e., disk-like bulges).
However, this is due to a bias caused by a priori selection of the $B/T$ range used for fitting the FJR.
Moreover, this selection leads to contradictory results because 
for brighter galaxies ($M_r\lesssim-20$ mag) a smaller $B/T$ corresponds to a larger $\sigma$  (see their Fig. 9).


\section{Discussion \label{sec:discussion}}

\begin{figure*}[t!]
\centering
\includegraphics[width=17cm]{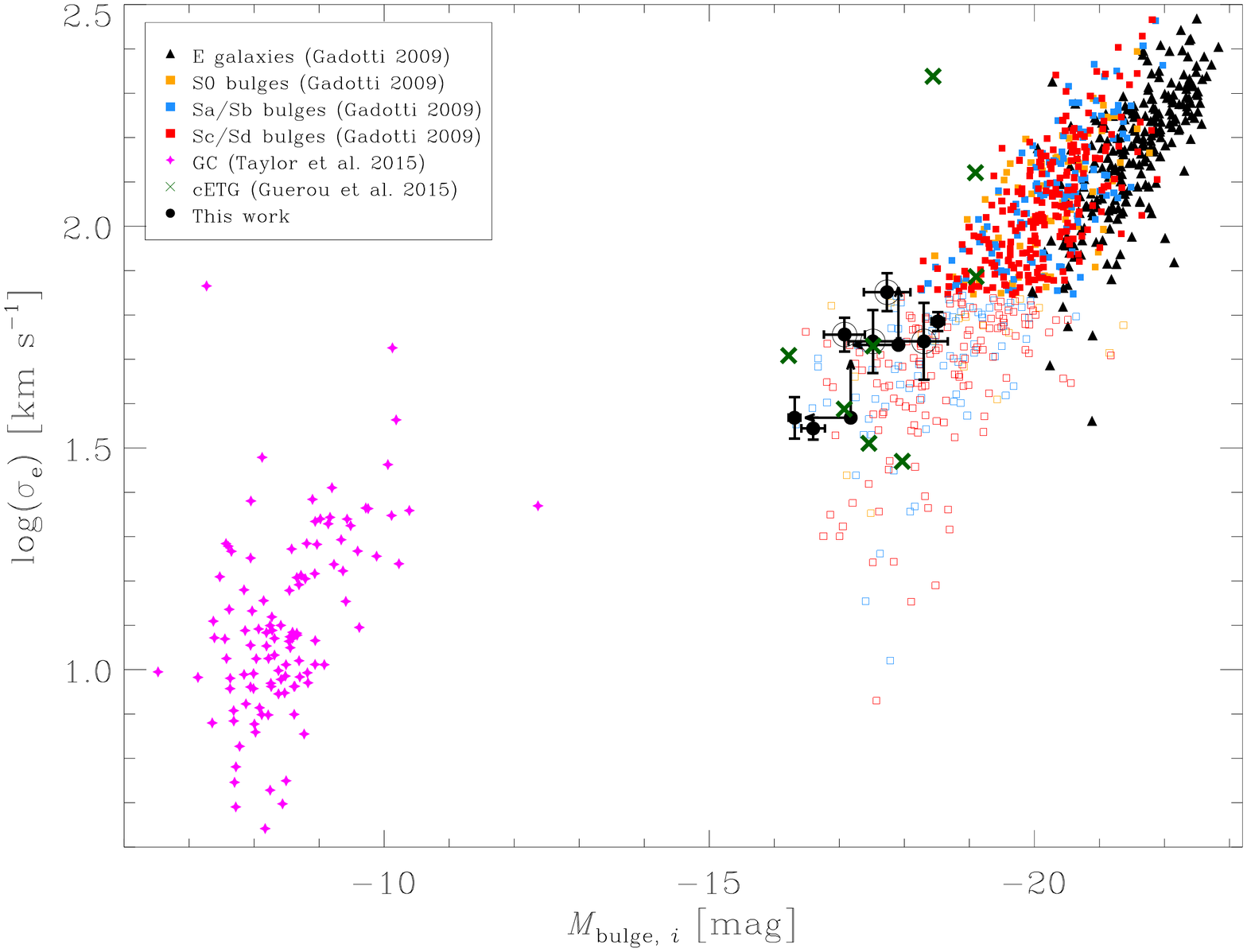}
\caption{As in Fig.~\ref{fig:faber_jackson} but including the GCs from
\citet[][pink stars]{Taylor2015} and cETGs from \citet[][green
crosses]{Guerou2015}.}
\label{fig:faber_jackson_total}
\end{figure*}

The 7 late-type bulges we studied in this work are small and have a
low mass, as confirmed by their position in the FPR
(Fig.~\ref{fig:fundamental_plane}), KR (Fig.~\ref{fig:kormendy}), and
FJR (Fig.~\ref{fig:faber_jackson}). Although they are located at the
low-$r_{\rm e}$ and low-$\sigma$ ends of the scaling
relations, they follow the same trend of ellipticals and larger and
more massive bulges. 
To address the statistical significance of our claim, we computed the 
probability of each bulge to be compatible with the fitted relations by 
means of MC simulations. The deviation of each bulge from the FPR 
and FJR is less than 2.1 rms, therefore we can not consider them as outliers.
We found no differences between photometric
and kinematic properties of barred and unbarred galaxies.
The remaining two galaxies SDSSJ1040 and SDSSJ1121 were 
better fitted only with a disk component. 
This means that either they have no bulge or their bulge 
is really small and beyond our possibility to detect it.
The latter reinforces the finding that small bulges are not low-$\sigma$ outliers in FJR.

As far as the structure and kinematics of disk-like bulges concerns,
they are expected to share the properties of the surrounding host
disks and be more rotation-dominated than classical bulges. 
Our 7 bulges fulfill many of the observational prescriptions
originally provided by \citet{Kormendy2004} and later revised by
\citet{Kormendy2016} for being classified as disk-like bulges.  They
have a late-type morphology with a nearly exponential light profile,
small $B/T$, low velocity dispersion, and ongoing star formation, as
proved by the presence of strong H$\beta$ and [\ion{O}{III}]5007
emission lines in their spectra.
All the galaxies of our sample exhibit either a spiral structure all the way
down to the galaxy center or a bar component
(Fig. \ref{fig:decomposition}). In this case the bulge dynamics should
be more similar to that of a disk rather than of a
spheroidal component, making the bulge a low-$\sigma$ outlier in the FJR. 
But our bulges are not rotation-dominated
systems ($|v_{\rm max}| \le 50$ km s$^{-1}$), although they are
characterized by very low values of velocity dispersion ($\sigma_{\rm  e} 
\la 70$ km s$^{-1}$). Each of them follows the same scaling
relations of ellipticals, massive bulges, and compact early-type
galaxies so they cannot be classified as disk-like systems.

We explored the possibility that our bulges could actually be similar
to other spheroidal systems, such as globular clusters (GCs) and
compact ETGs (cETGs). To this aim, we further extended the $M_i$ and
$\sigma_{\rm e}$ ranges of the FJR by including 125 GCs of the nearby
giant elliptical NGC~5128 \citep{Taylor2015} and 8 low-mass cETGs of
the Virgo cluster \citep{Guerou2015}. We transformed the absolute
magnitude of the GCs in the \emph{i} band using the prescriptions from
SDSS-DR8\footnote{The equations are available in 
\url{https://www.sdss3.org/dr8/algorithms/sdssUBVRITransform.php#Lupton2005}.}
\citep{Aihara2011}. 
Fig.~\ref{fig:faber_jackson_total} shows that our
bulges do not share the same position of GCs in the FJR despite they
have similar values of velocity dispersion ($\sigma_{\rm e}\simeq30$ km
s$^{-1}$). The absolute magnitude of GCs is indeed several orders of
magnitude fainter than that of small bulges. On the contrary, cETGs
follow the same FJR as our bulges and more massive ones. This supports the
idea that the low SDSS-based values of velocity dispersion are in
reality affected by instrumental resolution and disk contamination,
and suggests that the less massive bulges follow the same scaling
relations as ellipticals and more massive bulges.

We found no correlation between the location of our bulges in the FPR
and FJR and Hubble type of their host galaxies. The lack of
correlation between the structural and kinematical properties of
bulges and galaxy morphology is in agreement with the findings of
\citet{Thomas2006} who focussed on the stellar population properties
of bulges. The mass rather than the morphology seems to be the driver
of the intrinsic physical properties of bulges. As a matter of facts,
we found a single population of galaxy spheroids that follow the
same scaling relations, where the mass leads to a smooth
transition in the photometric and kinematic properties from less
massive bulges to more massive bulges and ellipticals.

It is worth noticing that classical and disk-like bulge are usually
separated only using the S\'ersic index of their surface-brightness
radial profile or even worse the S\'ersic index of their host
galaxy. This is a misuse of the findings of \citet{Fisher2008,
Fisher2010} who pointed out a bimodal distribution of the S\'ersic
index of bulges. Once bulge-disk decomposition was performed on
galaxies separated according to their visual morphology, they found
that classical bulges have $n>2$ while disk-like bulges have $n<2$.
To date, no physical explanation has been found for this behaviour.
Moreover, since the S\'ersic index correlates with $B/T$ in classical
bulges \citep{Fisher2008, Gadotti2009}, the common belief is that
disk-like bulges have $B/T < 0.35$. 
However, the photometric and kinematic properties of our bulges in the
framework of the scaling relations of ellipticals and bulges show that
small values of $n$ and $B/T$ do not guarantee that a bulge is disk
like.


\section{Conclusions \label{sec:conclusions}}

We analyzed the surface-brightness distribution and stellar kinematics
of a sample of late-type spirals. They were selected to investigate
the photometric and kinematic properties of small bulges in order to
understand whether they follow the same scaling relations traced by
ellipticals and large bulges and if they are disk-like or classical
bulges.

We obtained the structural parameters of the sample galaxies by
performing a two-dimensional photometric decomposition of their SDSS
\emph{i}-band images. The surface-brightness distribution of each
galaxy was assumed to be the sum of the contribution of a S\'ersic
bulge and an exponential disk. We included a Ferrers bar in fitting
the images of 4 sample galaxies. We found a bulge component in 7
galaxies, while the remaining 2 resulted to be pure disk galaxies. We
measured the stellar velocity dispersion within the bulge effective
radius from long-slit spectra taken with high spectral resolution.
All the sample bulges have small $r_{\rm e}$, nearly exponential light
profiles, small $B/T$, low $\sigma$, and ongoing star formation.

We combined the photometric ($r_{\rm e}$, $\langle \mu_{\rm e}
\rangle$, and $M_i$) and kinematic parameters ($\sigma_{\rm e}$) of
the sample bulges to study their location in the FPR, KR, and FJR. To
this aim, we built the scaling relations defined for a comparison
sample of nearby ellipticals and bulges with structural parameters and
velocity dispersions measured from SDSS \emph{i}-band images and
spectra by \citet{Gadotti2009} and \citet{Oh2011}, respectively. Our
data extend the scaling relations to the regime of bulges with $r_{\rm
  e} \simeq 0.2$ kpc, $\sigma_{\rm e}\simeq35$ km s$^{-1}$, and
$M_i\simeq-16$ mag.
The FPR coefficients are consistent with the findings of
\citet{Bernardi2003b}, which are based on a much larger sample of ETGs
and characterized by a smaller scatter. The location of the
ellipticals and bulges in the FPR suggests that there is a single
population of galaxy spheroids sharing the same physical properties.
This is also true when our small bulges are considered. No differences
appear in the residuals when galaxies are divided according to their
morphological type.
The sample bulges have actually small size and mass, as it results
from their location in the KR. We confirmed that this relation is a
poor proxy of the bulge properties and its large intrinsic scatter and
magnitude bias \citep{NigocheNetro2008, NigocheNetro2010}, 
although it is commonly used to disentangle between classical and
disk-like bulges \citep[e.g., ][]{Vaghmare2013, Mishra2017}.
Our small bulges with resolved $\sigma$ are not following the
downbending previously reported in the low-$\sigma$ end of the FJR, but trace the
same trend as ellipticals and large bulges. This supports the idea
that the lowest values of the SDSS-based $\sigma$ are actually
affected by instrumental resolution and disk contamination.  We
observed a smooth transition in FJR from larger to smaller values of
$n$ according to both $\sigma_{\rm e}$ and $M_i$ when the comparison
sample is divided in terms of the S\'ersic index. This suggests that
$n$ strongly depends on the mass of the system rather than on the
morphology of the host galaxy.

Our bulges fulfill most of the observational prescriptions
\citep{Kormendy2004, Kormendy2016} for being disk-like bulges. But,
they are not rotation dominated and follow the same scaling relations
of ellipticals and larger and more massive bulges, where the mass
seems to lead to a smooth transition in the photometric and kinematic
properties from less to more massive systems. We conclude that small
values of $n$ and $B/T$ do not guarantee that a bulge is disk like and not classical.

\begin{acknowledgements}

We would like to thank Mauro D'Onofrio for his comments 
on the FPR and the referee, Elmo Tempel, for his suggestions that 
helped us to improve the way we presented our results.
\\
This work is supported by Padua
University through grants 60A02-5857/13, 60A02-5833/14, 60A02-4434/15,
CPDA133894. E.~M.~C. and L.~M. acknowledge financial support from
Padua University grants BIRD164402/16 an CPS0204,
respectively. J.~M.~A. acknowledges support from the European Research
Council Starting Grant (SEDmorph; P.I. V. Wild).
L.~C. acknowledges the School of Physics and Astronomy
of the University of St. Andrews for hospitality while this paper was in progress.
This research also made use of the HyperLeda Database (\url{http://leda.univ-lyon1.fr/}) 
and NASA/IPAC Extragalactic Database (NED) which is operated by the 
Jet Propulsion Laboratory, California Institute of Technology, 
under contract with the National Aeronautics and Space Administration (\url{http://ned.ipac.caltech.edu/})
We acknowledge the use of SDSS data (\url{http://www.sdss.org}).

\end{acknowledgements}



\clearpage

\appendix

\section{Spatial resolution of the bulge component \label{appendix:resolution}}

\begin{figure}[h!]
\centering
\resizebox{\hsize}{!}{\includegraphics[]{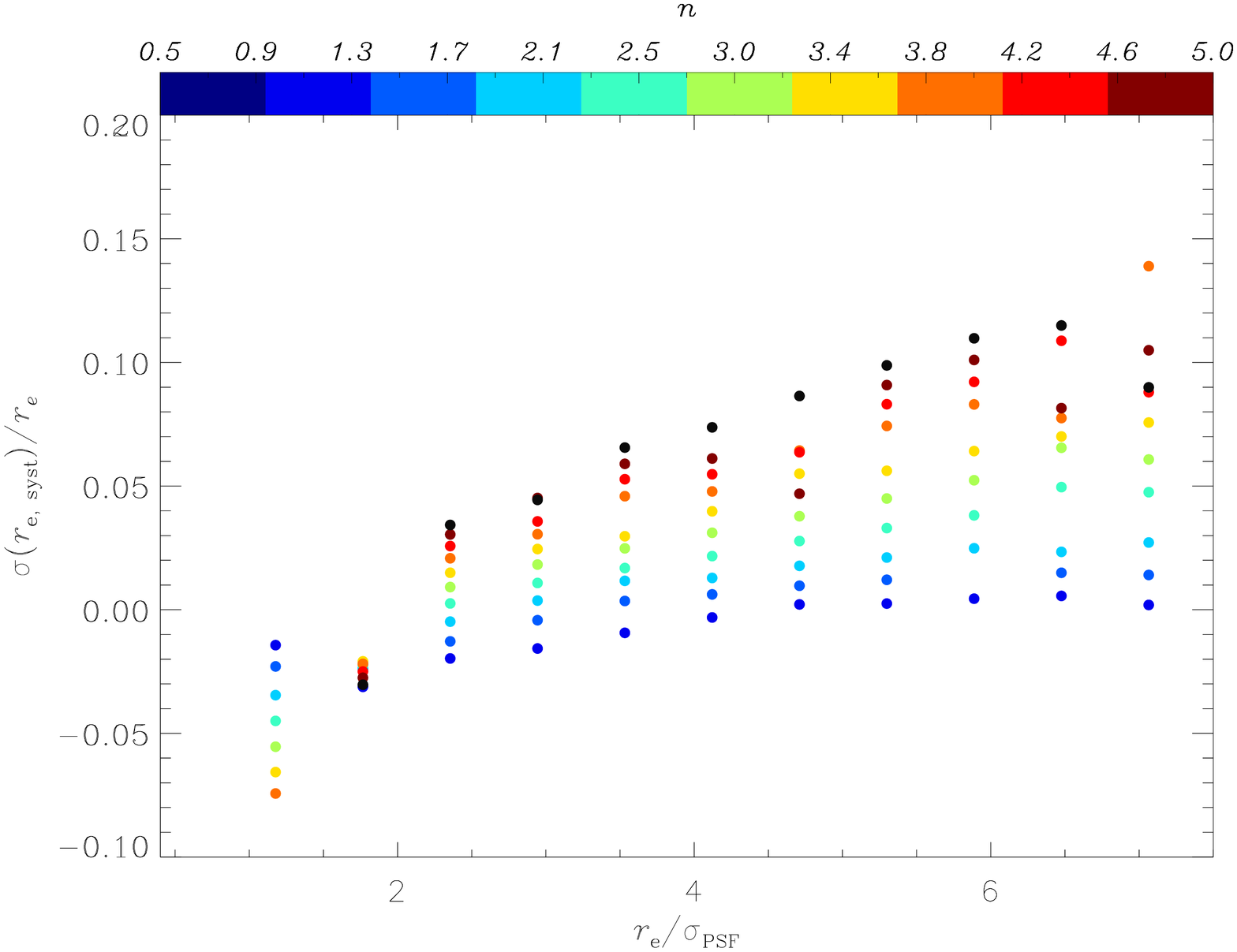}}
\caption{Relative difference between $r_{\rm e, \, output}$ and
$r_{\rm e, \, input}$ as function of $r_{\rm e}/\sigma_{\rm PSF}$ 
adopting $B/T=0.1$. Positive/negative values correspond 
to an over/underestimation of $r_{\rm e}$.} 
The different values of $n$ are shown in different colors.
\label{fig:resolution_re}
\end{figure}

\begin{figure}[h!]
\centering
\resizebox{\hsize}{!}{\includegraphics[]{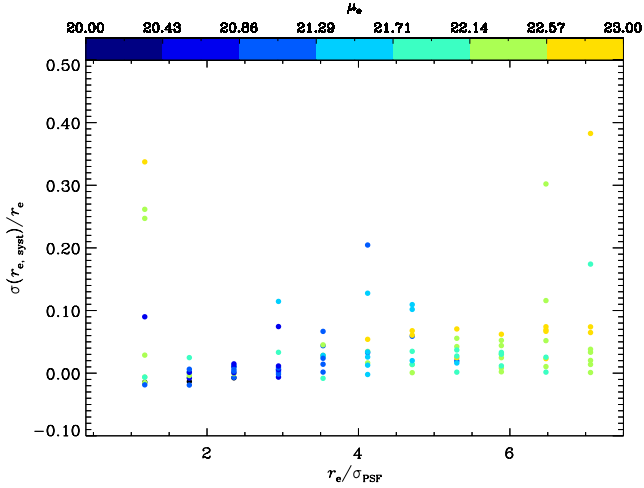}}
\caption{Relative difference between $r_{\rm e, \, output}$ and
$r_{\rm e, \, input}$ as function of $r_{\rm e}/\sigma_{\rm
PSF}$ adopting $h=5.6$ arcsec. Positive/negative values correspond 
to an over/underestimation of $r_{\rm e}$. The different values of $\mu_{\rm e}$ 
are shown in different colors.}
\label{fig:resolution_mue}
\end{figure}

We performed a further analysis to ensure that our bulges are
spatially resolved because the angular size of most of them is close
to the size of the image PSF ($1.3 < r_{\rm e}/\sigma_{\rm PSF} < 6.9$), as it results from
the photometric decomposition (Table~\ref{tab:decomposition}).

To this aim, we built a large number of images of mock galaxies with a
S\'ersic bulge and an exponential disk belonging to the faintest
magnitude bin ($m_i = 15$ mag), having a Moffat PSF with $\rm
FWHM_{\rm PSF}=1$ arcsec consistent with observations, and being
characterized by different values of $r_{\rm e}$ so that 
$r_{\rm e}/\sigma_{\rm PSF} = [1.2, 1.8, 2.4, 3.0, 3.5, 4.1, 4.7, 5.3, 5.9,
6.5, 7.1]$.  We chose the value of the S\'ersic index to be $n =
[0.5, 1, 1.5, 2, 2.5, 3, 3.5, 4, 4.5, 5]$.  We generated the remaining
structural parameters of the mock galaxies in two different
ways. First, we adopted $B/T = 0.1$ and derived $h$ using the $(r_{\rm e}/h, B/T)$ 
correlation given in Eq.~\ref{eq:rh_BT} to inspect
galaxies with different disk size. Then, we adopted $h=5.6$ arcsec
consistent with the disk scalelengths of our sample galaxies (Table
\ref{tab:decomposition}) and randomly derived $B/T$ from a uniform
distribution in the range $[0,0.45]$ to examine the impact of the
bulge component. We investigated all the possible permutations and
built two final samples of 110 mock galaxies each, where all the
geometrical parameters are the same for all the galaxies in the same
sample. We analyzed the images of the mock galaxies as if they were
real using GASP2D.

The two different sets of mock galaxies lead to the same results
(Fig.~\ref{fig:resolution_re} and \ref{fig:resolution_mue}). The bulge
component can be actually measured even if $r_{\rm e}/\sigma_{\rm PSF}
= $ 1.2 with a relative error $\sigma_{r_{\rm e},\,{\rm syst}}/r_{\rm e}\la0.1$ 
in agreement with the photometric errors from the MC
analysis. If brighter galaxies are considered (e.g., with $m_i = 13$
mag) the errors on $r_{\rm e}$ become smaller ($\sigma_{r_{\rm e},\,{\rm syst}}/r_{\rm e}\sim0.01$). 
Therefore, all the sample bulges are
spatially resolved. We also considered the role of $n$
(Fig.~\ref{fig:resolution_re}) an $\mu_{\rm e}$
(Fig.~\ref{fig:resolution_mue}). Galaxies with smaller $n$ are better
fitted than those with larger $n$. Moreover, galaxies with larger
$\mu_{\rm e}$ show a larger scatter around the actual value.  This
combined effect could be explained by the fact that smaller bulges are
also brighter and so their parameters are retrieved with more
accuracy.

\end{document}